\documentclass[journal,12pt,onecolumn,draftclsnofoot,]{IEEEtran}
\usepackage{cite}
\usepackage{amsmath,amssymb,amsfonts}
\usepackage{graphicx}
\usepackage{textcomp}
\usepackage{xcolor}
\usepackage{comment}
\usepackage{optidef}
\usepackage[T1]{fontenc}
\usepackage{mathtools} 
\usepackage{cuted}
\usepackage{bbm}
\usepackage{dsfont}
\usepackage{mathtools}
\usepackage{textgreek}
\usepackage{nicefrac}
\newcommand{\diag}{\operatorname{diag}}

\DeclareMathOperator{\Bernoulli}{Bernoulli}
\usepackage{subcaption}
\usepackage[linesnumbered,ruled,vlined]{algorithm2e}
\SetKwInput{KwInput}{Input}                
\SetKwInput{KwOutput}{Output}  

\DeclareCaptionFont{black}{\color{black}}

\usepackage{tabularx}
\newcounter{protocol}
\newenvironment{protocol}[1]
  {\par\addvspace{\topsep}
   \noindent
   \tabularx{\linewidth}{@{} X @{}}
    \hline
    \refstepcounter{protocol}\textbf{Protocol \theprotocol} #1 \\
    \hline}
  { \\
    \hline
   \endtabularx
   \par\addvspace{\topsep}}
\usepackage{amsthm}
\newtheorem{theorem}{Theorem}
\newtheorem{lemma}{Lemma}
\newcommand*{\Comb}[2]{{}^{#1}C_{#2}}%

\ifCLASSINFOpdf
\else
\fi

\begin{document}
%
\title{\LARGE Deep Learning-Based Blind Multiple User Detection for Grant-free SCMA and MUSA Systems}
%
%
%
\author{Thushan~Sivalingam,~\IEEEmembership{Graduate~Student~Member,~IEEE,} Samad~Ali, \IEEEmembership{Member,~IEEE,} Nurul~Huda~Mahmood, Nandana~Rajatheva, \IEEEmembership{Senior~Member,~IEEE,} and~Matti~Latva-Aho,~\IEEEmembership{Senior~Member,~IEEE}
\thanks{This research has been supported by the Academy of Finland, 6G Flagship program under Grant 346208. Also, this work is partially funded by the Hexa-X-II project receiving funding from the Smart Networks and Services Joint Undertaking (SNS JU) under the European Union’s Horizon Europe research and innovation programme under Grant Agreement No 101095759. Partial results from this article were presented in part at the IEEE PIMRC 2021~\cite{9569538,9569446}.}}
\maketitle
\vspace{-2cm}
\begin{abstract}
Massive machine-type communications (mMTC) in 6G requires supporting a massive number of devices with limited resources, posing challenges in efficient random access. Grant-free random access and uplink non-orthogonal multiple access (NOMA) are introduced to increase the overload factor and reduce transmission latency with signaling overhead in mMTC. Sparse code multiple access (SCMA) and Multi-user shared access (MUSA) are introduced as advanced code domain NOMA schemes. In grant-free NOMA, machine-type devices (MTD) transmit information to the base station (BS) without a grant, creating a challenging task for the BS to identify the active MTD among all potential active devices. In this paper, a novel pre-activated residual neural network-based multi-user detection (MUD) scheme for the grant-free SCMA and MUSA system in an mMTC uplink framework is proposed to jointly identify the number of active MTDs and their respective messages in the received signal's sparsity and the active MTDs in the absence of channel state information. A novel residual unit designed to learn the properties of multi-dimensional SCMA codebooks, MUSA spreading sequences, and corresponding combinations of active devices with diverse settings. The proposed scheme learns from the labeled dataset of the received signal and identifies the active MTDs from the received signal without any prior knowledge of the device sparsity level. A calibration curve is evaluated to verify the model's calibration. The application of the proposed MUD scheme is investigated in an indoor factory setting using four different mmWave channel models. Numerical results show that when the number of active MTDs in the system is large, the proposed MUD has a signiﬁcantly higher probability of detection compared to existing approaches over the signal-to-noise ratio range of interest.
\end{abstract}
\begin{IEEEkeywords}
DNN, grant-free, machine-type communications, multi-label classification, multi-user detection, MUSA, NOMA, ResNet, SCMA.
\end{IEEEkeywords}
%
\IEEEpeerreviewmaketitle
\section{Introduction}
%
%
%
%

\IEEEPARstart{M}{a}ssive machine-type communications (mMTC) is foreseen as one of the leading service classes for sixth-generation (6G) wireless communication systems~\cite{BroadbandWhitePaper2020}. mMTC supports various industrial, medical, commercial, defense, and general public applications in the Internet-of-Things (IoT) domain. mMTC focuses on supporting the uplink-dominated massive number of low-power and low-complexity devices that sporadically transmit short data packets~\cite{8663999} with low transmission rates during the short period of the active state. Therefore the conventional four-step scheduling-based multiple access schemes in which the base station (BS) allocates orthogonal time/frequency resources to each device are inefficient because of the heavy signaling overhead and higher access delay~\cite{6525600}.

\subsection{State-of-the-art}
Grant-free non-orthogonal multiple access (NOMA) schemes have been proposed as a promising solution to overcome the limitations mentioned above~\cite{7263349}. Grant-free access allows the machine-type devices (MTDs) to communicate pilot and information symbols without acquiring a transmission grant, significantly reducing the signaling overhead. NOMA techniques have been extensively investigated to potentially support a massive number of MTDs by sharing a limited amount of time and frequency resources in a non-orthogonal manner~\cite{7263349}. In this approach, there is inter-user interference because of the orthogonality violation. Therefore, NOMA applies device-specific non-orthogonality sequences to mitigate the inter-user interference~\cite{7263349}. Several signature-based NOMA schemes were proposed based on device-specific codebook structures, interleaving patterns, delay patterns, scrambling sequences, and spreading sequences~\cite{8486939}. From this perspective, the sparse coded multiple access (SCMA) is designed as a code-domain-NOMA technique~\cite{6966170} by Huawei, and multi-user shared access (MUSA) is introduced by ZTE~\cite{7504361} as a spreading-sequence-based NOMA scheme.

Recognizing the underlying challenge of grant-free access that each MTD transmits information without scheduling, a mechanism is required at the BS to detect the active MTDs among all inherent devices in the network. This procedure is defined as multi-user detection (MUD). In mMTC network, a single MTD is not active for a long period, and only a few devices are simultaneously active. Due to the infrequent nature of mMTC traffic, the activity vector, which marks the active devices out of all available MTDs, can be represented as a sparse vector. This problem can be formulated as a sparse signal recovery problem by considering the sparsity of the activity vector. Furthermore, based on the antenna configuration of the BS, this problem can be regarded as a single measurement vector (SMV) MUD problem where BS has a single antenna or a multiple measurement vector (MMV) MUD problem where BS has multiple antennas. Solving this problem involves two strong subqueries, i) how many MTDs are active (sparsity) and ii) which are those active MTDs, where the BS strictly does not have complete or partial knowledge of sparsity under grant-free NOMA. 

A number of studies based on compressive sensing (CS) theory have recently been proposed to exploit the sparse characteristic of device activity~\cite{7462187,7976275,7551125}. The study~\cite{7462187} investigates the structured sparsity of the active users and proposes a low-complex structured CS-based iterative algorithm to detect the active users and data jointly. To improve the MUD performance, the authors in~\cite{7976275} propose a prior-information-aided adaptive subspace pursuit (PIA-ASP) algorithm to exploit the intrinsically temporal correlation of active user support sets in several continuous-time slots. Moreover, dynamic CS-based MUD proposed in~\cite{7551125} to detect both active users and their corresponding data jointly in several continuous-time slots by exploring the temporal correlation of the active user sets. The authors~\cite{8937497} propose a joint MUD and channel estimation, focusing on reducing the pilot length and computational complexity using dimension reduction. Another study~\cite{8675520} investigates an expectation propagation-based scheme for MUD using the Bernoulli-Gaussian probabilistic model. In addition, the designers of SCMA and MUSA started the ball rolling on blind MUD~\cite{6933472,8288402}. Reference~\cite{6933472} proposes an iterative algorithm to detect the active pilots and decoding mechanism of user's data. In~\cite{8288402}, the authors present MUD without the reference signal at the BS for MUSA. First, the spreading codes used by the active users are estimated by an iterative algorithm. Then the active users are detected by blind equalization. 

In a high overloading ratio (OR), where more devices are active than the available resources, the performance of CS-based approaches~\cite{7462187,7976275,7551125,8937497,8675520,6933472,8288402} degrade due to the increased correlation between the columns of the sensing matrix. The increased sparsity of the input vector also influences the performance negatively. Besides, conventional CS-based solutions strongly depend on channel estimation quality. However, perfect channel estimation cannot be accomplished in practical mMTC. The aforementioned iterative algorithms also take considerable time to converge, increasing the communication latency. Even though a significant amount of literature has been published on conventional algorithms, the literature has not thoroughly addressed the realistic nature of the MUD for mMTC. Taken together, developing a practically feasible, scalable MUD scheme for grant-free code domain NOMA (SCMA and MUSA) for mMTC is a challenging open problem.

Cutting-edge explorations and expansions of machine learning have started to address critical problems in wireless communication driven by the advancement of computational capabilities and algorithm complexities~\cite{6GFlagship_WP2,9535455}. The authors~\cite{9535455} discuss the challenges of deploying NOMA in 5G and beyond for mMTC, such as non-unified signal processing architectures of various NOMA schemes and the inflexibility resulting from the offline design paradigm. Also suggests that deep learning can be used to overcome these challenges by constructing a deep neural networks (DNN) with a uniform signal-processing architecture. Few studies have attempted to explore the MUD problem by deep learning in mMTC. Two studies~\cite{8968401,9149252} propose MUD for a grant-free low-density signature (LDS) scheme. In~\cite{8968401}, the authors present deep learning-based parallel receivers with a softmax estimator for each distinct sparsity level. However, sparsity estimation based on softmax thresholding performs better only for specific sparsity levels where the pre-defined threshold value satisfies the estimation conditions. Designing a distinct MUD strategy for each sparsity level is also impractical and limits the algorithm's scalability. From a theoretical point of view, softmax takes the output value of the layer before and produces a probability distribution, where output values are interrelated. However, still, the activities of MTDs are not correlated. Also, ensemble learning has a risk of carrying a high bias toward its aggregate and is undoubtedly expensive in computational and implementation complexity. In the study~\cite{9149252}, the authors set the number of active devices during the training, which does not enable DNN to learn the entire codebook of the network. Therefore, it induces misdetection during practical implementation. In~\cite{9432908}, the authors propose a joint channel estimation and device detection for a massive multiple input multiple outputs (MIMO) system using unique pilot sequences. The device activity assumes a Bernoulli-Gaussian mixture, an implicit sparsity assumption. Also, the judgment threshold-based hard decision limits the algorithm's scalability. A recent study on joint MUD and channel estimation for unique spreading sequences~\cite{9605579} investigate the MUD with partially known support as prior knowledge. Likewise,~\cite{9530373} presents MUD and constellation design for grant-free NOMA with channel state information (CSI) at the decoder for a small group (four devices and six resources) of devices. Another MUD algorithm is presented in~\cite{9789271} based on Zadoff-Chu using a hard decision threshold compared with Bernoulli and normal sequences. Furthermore, using the LDS codebook, a convolutional neural network (CNN)~\cite{9462894} presents MUD for MMV systems under a specific sparsity level and antenna count. In~\cite{8952876}, the authors propose a multi-task deep neural network framework called DeepNOMA to optimize NOMA for IoT. It includes an auto-encoding structure channel module, a multiple access signature mapping module (DeepMAS), and a multi-user detection module (DeepMUD) trained in a data-driven fashion. In addition, the authors of~\cite{9697978} propose a blind signature classification for NOMA using a deep learning framework called DeepClassifier, which achieves classification accuracy with reduced computational complexity by utilizing a memory-extended structure and a deep recurrent accumulation neural network. Furthermore, the work in~\cite{8625480} discusses the challenges of achieving ultra-responsive and ultra-reliable connections for massive IoT devices in the context of tactile IoT. The authors propose a nonlinear spreading sequence and an end-to-end neural network model for grant-free NOMA using deep variational auto-encoder to solve the optimization problem, which includes random user activation and symbol spreading.

Even though recent studies have started to explore MUD using deep learning, a significant amount of solid aspects have not been addressed. Significantly, perfect CSI at the receiver, complete or partial knowledge of sparsity, and threshold-based decision-making at receiver limit the scalability and perfectness of MUD. Furthermore, only a few studies considered specific deep learning parameters (recall, precision, area under the curve (AUC)) to validate the performance; still, none of the work explored calibration curves to validate the model, which is a robust evaluation method for classification problem~\cite{10.5555/3305381.3305518}. Again, only LDS codebook investigation is available in the code domain NOMA, which cannot be directly applicable to SCMA and MUSA.

\subsection{Challenges}
This research focuses on investigating the five main challenges associated with MUD in mMTC scenarios. The first challenge is to develop a DNN-based MUD model that can handle various code-domain NOMA schemes (SCMA and MUSA) effectively without having prior knowledge of sparsity and CSI. The second challenge is related to the code-domain schemes, where limited codebooks are available for SCMA techniques, supporting up to six devices only, which is insufficient for mMTC scenarios. In addition, there is currently no spreading sequence selection algorithm established for MUSA techniques. Third, evaluating both SMV and MMV systems is crucial for a comprehensive understanding of MUD in mMTC scenarios. Specifically, studying SMV for various channel models is essential. The fourth challenge is designing an optimized deep-learning model with the appropriate internal parameters, which presents two main difficulties, such as the unavailability of a suitable dataset and a readily available deep-learning framework. Finally, proper calibration of the deep learning model (classification) prior to evaluation is crucial but has received limited attention in the field of wireless communication. This research aims to address these five challenges to provide insights into the development of advanced MUD algorithms for mMTC systems.

\subsection{Contribution}
The major contributions of this study are as follows:
\begin{itemize}
   \item{\textcolor{black}{Our proposition entails a novel pre-activated residual neural network (ResNet)-driven blind DNN-MUD architecture that includes a novel residual unit with diverse settings, addressing grant-free NOMA schemes such as SCMA codebooks and MUSA spreading sequences with various overload ratios. Our proposed architecture enables the learning of the correlation between the received signal and the SCMA codebooks and MUSA spreading sequences with the combination of device sparsity levels in the signal-to-noise ratio (SNR) range during the offline training. Furthermore, the proposed architecture jointly detects the received signal's sparsity and corresponding active devices using a sigmoid estimation online without CSI.}}
    \item{\textcolor{black}{We formulate an SMV sparse signal recovery problem for jointly achieving sparsity and active devices. In addition, we re-structure the received measurement vector by separating the real and imaginary components and stacking them together as an input vector to form the training data. Alongside this, we annotate the MTDs' statuses to produce the training labels. The DNN is then fed with the aforementioned training data and labels to facilitate the learning of internal parameters, which include entries for the codebook and spreading sequences, as well as combinations of active devices in the mMTC system.}}
    \item{\textcolor{black}{In view of the unavailability of readily accessible SCMA codebooks and MUSA spreading sequences for massive devices, we undertake the design of SCMA codebooks for different ORs through the utilization of constellation rotation and interleaving techniques as prescribed in~\cite{6966170, 7504356} and generate a feasible set of MUSA spreading sequences based on~\cite{7504361}. However, the task of selecting the best MUSA spreading sequences from the available set poses a challenge. Consequently, we propose a scalable heuristic algorithm aimed at identifying the spreading sequences with the desired orthogonality factor from the potential set. This is undertaken to minimize the correlation between the MUSA spreading codes. In addition, we proposed a collisional avoidance protocol for MUSA schemes.}}
    \item{\textcolor{black}{We further expand our DNN algorithm to address the MUD of the MMV system, where we separate the real and imaginary elements of each antenna measurement, which are subsequently concatenated serially to construct the input training data. Since the sparsity is the same, we opt to employ the exact SMV methodology for annotating the training labels.}}
    \item {\textcolor{black}{We derive the computational complexity of the proposed DNN-MUD architecture and provide the complexity of the well-known algorithms previously referenced to evaluate the overall efficiency and determine the extent of their effectiveness compared to the proposed MUD. Also, show the theoretical investigation for labelset selection and convergence.}}
    \item{\textcolor{black}{We validate our model using a standard calibration curve and conduct a comprehensive assessment of the proposed DNN-MUD performance. This involves a comparative analysis with well-established algorithms, such as least squares-block orthogonal matching pursuit (LS-BOMP)~\cite{8570860}, complex-approximate message passing (C-AMP)~\cite{6478821}, stagewise orthogonal matching pursuit (stOMP)~\cite{6145475}, Deep-MUD (D-MUD)~\cite{8968401}, and CNN-MUD~\cite{9462894} algorithms. Furthermore, we evaluate the efficacy of the DNN-MUD algorithm in a practical setting by simulating the SMV scenario in an indoor factory environment operating at mmWave frequencies~\cite{3gpp.138.901}.}}
\end{itemize}

The structure of our paper is as follows. Section II presents the system model and problem formulation, including the concept, generation, and code selection of SCMA codebook and MUSA spreading sequences. Section III describes the deep learning approach and MUD structure for SMV and MMV systems. Section IV shows the complexity, training, and testing data generation. Section V presents the simulation setup, parameters, and numerical results. Finally, Section VI concludes the study. 
\subsubsection*{Notations}
Boldface uppercase, boldface lowercase, and lower case letters represent matrices, vectors, and scalars, respectively, whereas calligraphy letters denote sets. \begin{math} \mathbb{R} \end{math} and \begin{math} \mathbb{C} \end{math} denote the space of real and complex numbers, respectively. The operations $(.)^H$ and $(.)^T$ denote conjugate transpose and transpose, respectively. $\Re(s)$ and $\Im(s)$ are the real and imaginary part of a complex number $s$, respectively. $\mathbf{I}$ denotes the identity matrix, where the size is evident from the context. In addition, complex Gaussian distribution with zero mean, and variance ${\sigma_w^2}$ is represented by \begin{math} \mathcal{CN} (0,{\sigma_w^2}) \end{math}. The Hadamard (element-wise) product operator is denoted by $\circ$. Also, the absolute value of the complex number $x$ is represented by $|x|$, Euclidean norm of the vector $\mathbf{x}$ is denoted by ${\lVert\mathbf{x}\rVert}$, \textcolor{black}{and $\lfloor x \rfloor$ indicate round down $x$ to the nearest integer.}

\section{System Model and Problem Formulation}
\subsection{System Model}
\begin{figure}[t]
    \center
    \includegraphics[width=0.60\linewidth, height=0.60\textheight,keepaspectratio]{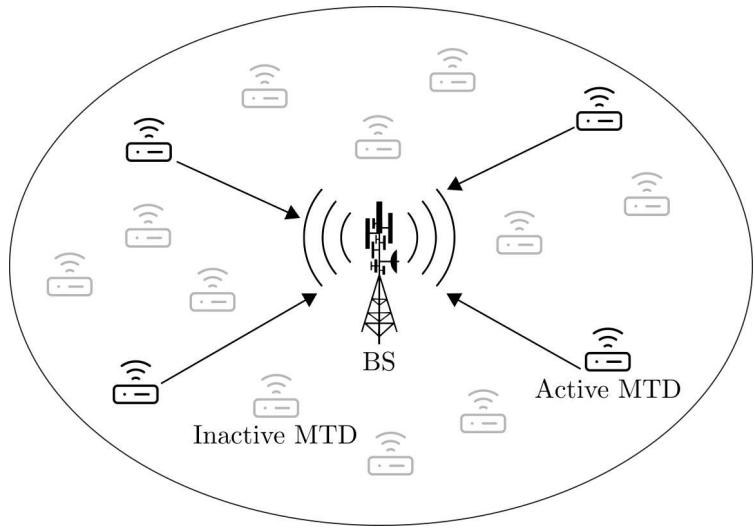}
    \caption{The illustration of mMTC grant-free NOMA uplink communication system where a fraction of devices transmits information to the BS.}
    \label{fig:System Model}
\end{figure}

Consider the uplink grant-free NOMA system of a mMTC network where a set $\mathcal{N}$ of $N$ randomly distributed MTDs are served by a BS. In the SMV scenario, the BS and MTDs are each equipped with a single antenna, and in the MMV scenario, the BS is equipped with multiple antennas, and MTDs are equipped with a single antenna. In this study, we focus on the overloaded scenario, where the number of MTDs is higher than the available uplink radio resources $K$ \begin{math} (K < N) \end{math} and OR $\triangleq \nicefrac{N}{K}$. Each MTD operates independently in this network. Furthermore, only a small subset of MTDs is active $n$ out of $N$ at any given time; the rest of the MTDs are inactive, shown in Fig.~\ref{fig:System Model}. We consider a slotted time access mechanism where the active MTDs at a given time slot access the medium synchronized in time. All active MTDs communicate to the BS after spreading with the device-specific non-orthogonal codebook in SCMA and non-orthogonal complex spreading sequences in MUSA schemes. Since each active MTDs share information spontaneously without scheduling, the BS needs to classify the active MTDs, i.e., determine the number of active devices, their identity, and transmitted data. 
\subsection{Problem Formulation}
We define the binary state indicator (active or inactive) $a_i$ of the $i$-th MTD as
\begin{equation}
 {a_i} = 
\begin{cases}
1, \ $ $i$-th MTD is active,$\\
0, \ $ $i$-th MTD is inactive.$
\end{cases}
\end{equation}
The probability of the active state of the $i$-th MTD is $p_i$, which is independent of other devices in the cell. The received signal \begin{math} \mathbf{y} \in \mathbb{C}^K \end{math} at the BS can be expressed as
\begin{equation}
\mathbf{y} = \sum_{i=1}^{N}{a_i}{\mathbf{s}_i}{{h}_i}{x}_i + \mathbf{w},
\end{equation}
where \begin{math} {\mathbf{s}_i} \in \mathbb{C}^K \end{math} is the spreading sequence vector of the $i$-th MTD (device specific codeword vector for SCMA and complex spreading sequence for MUSA), $h_i$ is the complex uplink channel coefficient from the $i$-th MTD to the BS which includes both small and distance-dependent large scale fading, $x_i$ is the transmit symbol of the $i$-th MTD, and \begin{math} \mathbf{w} \in \mathbb{C}^K \sim \mathcal{CN} (0,{\sigma_w^2}\mathbf{I})\end{math} is the complex-Gaussian noise vector. As specified earlier, at the beginning of the transmission, BS does not have knowledge of the number of active devices and corresponding spreading sequences. Therefore BS performs MUD to identify them. In particular, all the active MTDs transmit the pilot symbol $x_{p,i}$ to the BS in order to assist it to detect the active devices and corresponding spreading sequences ($\mathbf{s}_i$). After that, it estimates the channel coefficients ($h_i$). Finally, BS decodes the $J$ data symbols $x_{d,i}^{[1]}, \dots, x_{d,i}^{[J]}$ transmitted by the active MTDs. The transmission is over $J+1$ slots, where $J$ is selected such that the channel coherence time is greater than the transmission time of $J+1$ transmission slots. The pilot measurement vector \begin{math} \mathbf{y}_{p} \in \mathbb{C}^K \end{math} is given by
\begin{equation}
\mathbf{y}_{p} = \sum_{i=1}^{N}{a_i}{\mathbf{s}_i}{{h}_i}{x}_{p,i} + \mathbf{w}.
\label{eq:generation}
\end{equation}
Let us define $\boldsymbol{\phi}_i$ = $\mathbf{s}_i$${x}_{p,i}$, and $\boldsymbol{\Phi}$ = [{$\boldsymbol{\phi}_1$}, \dots, {$\boldsymbol{\phi}_N$}], we then have 
\begin{equation}
    \mathbf{y}_{p} = \sum_{i=1}^{N}{\boldsymbol{\phi}_i}{a_i}{{h}_i} + \mathbf{w} = \boldsymbol{\Phi}(\boldsymbol{a} \circ \mathbf{h}) + \mathbf{w},
\label{eqyp}
\end{equation}
where $\boldsymbol{a}$ = $[a_1, \dots, a_N]^T$ is the activity vector, and \begin{math} \mathbf{h} \end{math} = $[{h}_1, \dots, {h}_N]^T$ is the channel vector. Let us define $\boldsymbol{\varphi} = (\boldsymbol{a} \circ \mathbf{h}) = [a_1h_1, \dots, a_Nh_N]^T$. We can rewrite (\ref{eqyp}) as
\begin{equation}
    \mathbf{y}_{p} = \boldsymbol{\Phi}\boldsymbol{\varphi} + \mathbf{w}.
 \label{eqfinal}   
\end{equation}
We consider a fraction of MTDs ($n$ out of $N$) to be active at any given time. The accumulated sparse vector $\boldsymbol{\varphi}$ has $n$ non zero elements. Therefore, the received vector $\mathbf{y}_{p}$ can be represented as a linear combination of $n$ submatrices of ${{\phi}_1 \dots {\phi}_N}$ perturbed by the noise. There is an essential difference in the formulation of this problem for both SCMA and MUSA schemes. The actual code assignments for the two systems are different from each other. In SCMA schemes, BS knows the device-specific non-orthogonal codebook for each MTD. That means $\boldsymbol{\Phi}$ is available at the BS. The main task of the BS is to identify the $n$ submatrices in the received vector, equal to the number of active devices. For example, if the first and the third devices are active, then $\boldsymbol{\phi}_1$ and $\boldsymbol{\phi}_3$ are the components of $\mathbf{y}_{p}$. In MUSA sequence-based NOMA schemes, each MTD can choose the spreading sequences independently, which eases the resource coordination of the BS. \textcolor{black}{Note that a single MTD is not active for a long period, and only a few devices are active in a particular time frame. Considering the infrequent nature of mMTC traffic, the probability of choosing the same spreading sequence out of the massive number of sequences for two or more devices is extremely low. In addition, we propose an approach to avoid collision in Protocol $1$.} Here, BS knows all the available spreading sequences. Therefore, BS has an identity for all the available codes. Consequently, the objective of MUD at the BS in MUSA scheme is to detect the $n$ number of codes and the code identity involved in the received signal. For instance, if the second and the third codes are taken by active MTDs, then $\boldsymbol{\phi}_2$ and $\boldsymbol{\phi}_3$ are the components of \begin{math} \mathbf{y}_{p} \end{math}. Hence, the MUD problem can be formulated as the sparse signal recovery problem:
\begin{equation}
       \Omega = \arg\min_{\mid\Omega\mid=n} \frac{1}{2} \parallel \mathbf{y}_p- \boldsymbol{\Phi}_\Omega\boldsymbol{\varphi}_\Omega \parallel_2^2.
       \label{argmin}
\end{equation}
\subsection{Codebook-SCMA}
There are no available codebooks for SCMA beyond six devices and no straightforward code generation approach. We address this issue by proposing a simple but efficient heuristic codebook generation procedure. SCMA is designed as a non-orthogonal multi-dimensional structured codebook with layer-specific shaping gain where shaping gain is achieved by utilizing the QAM modulation~\cite{6966170}. The fundamental approach of SCMA is directly mapping the incoming MTD's bits into a codeword vector, where bit-to-symbol mapping and spreading are coupled. Furthermore, SCMA codewords are sparse, whereas the non-zero entries in the codeword are the assigned orthogonal channel resources for the uplink communication to the specific MTD~\cite{8434102}. The number of non-zero elements in the codeword is defined as a diversity order of the codeword. Moreover, the same diversity order assigns to different devices. In this work, we design the codebooks based on~\cite{6966170,7504356}. The design includes constellation rotation and interleaving. The multi-dimensional SCMA codebook of $i$-th MTD device \begin{math} {\mathbf{S}_i} \in \mathbb{C}^{K \times R} \end{math} is defined as
\begin{equation}
    \mathbf{S}_i = \mathbf{V}_i\triangle_i\mathbf{M}_c,\ \text{for} \ i = 1, 2,\dots ,N,
    \label{scma}
\end{equation}
where $R$ is the constellation points, \begin{math} {\mathbf{V}_i} \in \mathbb{B}^{K \times U} \end{math} is the $U$-dimensional SCMA binary mapping matrix of $i$-th MTD, $\triangle_i$ is the constellation operator of the $i$-th MTD, and $\mathbf{M}_c$ is the mother constellation. The matrix ${\mathbf{V}_i}$ can be generated by including $K-U$ all-zero row vectors inside the rows of $\mathbf{I}_U$~\cite{6966170}. Furthermore, the SCMA design contains the properties below: $N = \Comb{K}{U}$ and number of MTDs connected to the same resource, $d_f = \Comb{K-1}{U-1}$.
 Details of $\triangle_i$ and $\mathbf{M}_c$ are described in the below subsections. The design of SCMA consists of two major parts, designing the multi-dimensional mother codebook and constructing the codebook based on the number of MTDs.
\subsubsection{Design of SCMA mother codebook}
First, two-dimensional lattice ($\mathbf{Z}^{2}$) with $R$ points is defined as the first dimension of the mother constellation, where the steps of the design and theory of the lattice constellation are described in~\cite{6966170}. Then, we perform the Gray mapping for each element in the first dimension set. After that, the remaining required number of dimensions is generated using the first dimension's rotation with the corresponding Gray mapping, which gives us the complete mother constellation. However, this constellation is not efficient enough 
to control the effect of channel fading~\cite{7504356}. Therefore, to improve the energy efficiency of the codeword and reduce the peak to average power ratio, the authors in~\cite{7504356} propose to use interleaving. Furthermore, they claimed that this interleaving could improve the performance in the fading channels. During interleaving, even rows of the initial constellation are reordered as given in~\cite{7504356} while the odd rows are kept as before, which provides us with the final mother constellation ($\mathbf{M}_c$).

\subsubsection{Generation of factor graph for a separate codebook}
The next step is to construct a device-specific codebook for each MTD. Two significant conditions must be followed while constructing the device-specific codeword, such as retaining the Euclidian distance profile and structure of the mother codebook. The authors in~\cite{6966170,7504356} propose to use Latin square and factor graph-based constellation to ensure the conditions. The factor graph representation describes the relation between the MTDs and the resource elements. The factor graph matrix of this system described as $\mathbf{F} = (\mathbf{f}_1, \dots, \mathbf{f}_N)$, where each vector $(\mathbf{f}_i)$ contains $1$s and $0$s. The elements of the vector, $(\mathbf{f}_i)$ assigned based on $U$, and $d_f$ where there are $U$ non-zero elements in each column and $d_f$ number of non-zero elements in each row in the $\mathbf{F}$ matrix. Factor graph matrix for $N = 8$, $K = 6$, $U = 3$, and $d_f = 4$ given for example as
\begin{equation}
    \mathbf{F} = \begin{bmatrix}
1 & 1 & 1 & 1 & 0 & 0 & 0 & 0\\
0 & 0 & 0 & 0 & 1 & 1 & 1 & 1\\
0 & 1 & 0 & 0 & 0 & 1 & 1 & 1\\
1 & 0 & 1 & 1 & 1 & 0 & 0 & 0\\
0 & 1 & 0 & 0 & 0 & 1 & 1 & 1\\
1 & 0 & 1 & 1 & 1 & 0 & 0 & 0
\end{bmatrix}.
\end{equation}
Next, we replace the $d_f$ non-zero elements of each row of the factor graph matrix $\mathbf{F}$ with each of the $d_f$ phase rotation angles $\varrho_i$ according to the Latin structure. This ensures that the any given phase rotation angle does not appear twice within the same row or column. The phase rotation angle $\varrho_i$ is defined as~\cite{5425243}
\begin{equation}
    \varrho_i = \exp{\left(j\frac{2\pi}{Rd_f}(i-1)\right)}, i = 1, 2,\dots, d_f.
\end{equation}
As an example, for $d_f$ = 4, the phase rotation angles will be $(\varrho_1, \varrho_2, \varrho_3, \varrho_4)$, and the corresponding updated factor graph matrix is
\begin{equation}
    \tilde{\mathbf{F}} = \begin{bmatrix}
\varrho_1 & \varrho_2 & \varrho_3 & \varrho_4 & 0 & 0 & 0 & 0\\
0 & 0 & 0 & 0 & \varrho_1 & \varrho_2 & \varrho_3 & \varrho_4\\
0 & \varrho_3 & 0 & 0 & 0 & \varrho_4 & \varrho_1 & \varrho_2\\
\varrho_4 & 0 & \varrho_1 & \varrho_2 & \varrho_3 & 0 & 0 & 0\\
0 & \varrho_4 & 0 & 0 & 0 & \varrho_1 & \varrho_2 & \varrho_3\\
\varrho_3 & 0 & \varrho_4 & \varrho_1 & \varrho_2 & 0 & 0 & 0
\end{bmatrix}.
\end{equation}
Finally, considering $\tilde{\mathbf{f}_i}$, the $i$-th column of the updated factor graph matrix $\tilde{\mathbf{F}}$ without zero entries; the device-specific constellation operator \begin{math} \triangle_i \in \mathbb{C}^{U \times U} \end{math} for $i$-th MTD can be define as
\begin{equation}
    \triangle_i = \diag(\tilde{\mathbf{f}_i}).
\end{equation}
\subsection{Spreading sequence-MUSA}
MUSA spreading sequences are designed to support massive MTDs in a dedicated amount of radio resources. There are no available MUSA spreading sequences with different ORs. Here, we explore the generation briefly. In the MUSA scheme, each MTD information is spread with a group of low cross-correlation complex spreading sequences. BS can generate a large set of possible MUSA codes with specific $M$-ary values and required code lengths (number of radio resources). Also, MUSA codes are designed to support high OR with small code lengths to efficiently handle power consumption and delay. Furthermore, the different $M$-ary \begin{math}(M \geq 2)\end{math} values and code length define the space of available MUSA sequences~\cite{7504361}. MUSA considers the independence of choosing real and imaginary parts of the complex spreading elements; we consider $M = 3$ codes for our study, which is recommended by the MUSA designers~\cite{7504361} and able to produce an adequate amount of codes. $M = 3$ codes are generated from the basic element set $\{-1, 0, 1\}$ and the $9$ complex spreading elements are generated from the basic element set as $\{-1+i,-1,-1-i,i,0,-i,1+i,1,1-i\}$. We can create the space of all possible MUSA sequences from the complex spreading elements by permuting them according to the required code length. Therefore, the set \begin{math} \mathcal{M} \end{math} of $9^K$ complex sequences can be generated from $M = 3$ codes. However, we need only $N$ number of MUSA spreading sequences. Out of all available sequences, some are highly cross-correlated sequences, which reduces the MUD performance. Rather than randomly choosing $N$ columns of all possible MUSA sequences, we propose a scalable heuristic algorithm to choose a required number of low-cross correlated sequences, summarised in Algorithm~\ref{Alg:codeselection}. Given a cross-correlation threshold $\rho$, the proposed algorithm selects a subset \begin{math} \tilde{\mathcal{M}} \end{math} of \begin{math} \mathcal{M} \end{math}, such that the mutual cross-correlation between any two elements of \begin{math} \tilde{\mathcal{M}} \end{math} is less than or equal to $\rho$.
\begin{algorithm}[tb]
  
  \KwInput{Cross-correlation threshold $\rho$}
  \KwOutput{A matrix $\tilde{\mathcal{M}}$ of MUSA sequences with low mutual cross-correlation}
  \KwData{Matirx ${\mathcal{M}}$ with all possible combination of MUSA sequences}
  \textbf{Initialization:} Candidate columns $\mathcal{C} = \mathcal{M}; \tilde{\mathcal{M}} = \emptyset $\;
\While{$ \mathcal{C} $}
   {
           $\mathbf{m}$ = randomly selected column from $\mathcal{C}$\;
           $\tilde{\mathcal{M}} = \tilde{\mathcal{M}} \cup \mathbf{m}$\;
           $\mathcal{C} = \left\{ \mathbf{c} \mid \mathbf{c} \in \mathcal{C}, \left(\frac{\mathbf{c}}{\|\mathbf{c}\|}\right)^T\left(\frac{\mathbf{m}}{\|\mathbf{m}\|}\right) \leq \rho \right\}$
          
   }
\caption{A heuristic algorithm to select low-cross correlated MUSA sequences}
\label{Alg:codeselection}
\end{algorithm}

\textcolor{black}{In mMTC, few MTDs are high-active, and other MTDs are low-active. We assume to have the information based on statistics. Therefore, we define the number of high-active MTDs as $N_{high}$ and low-active MTDs as $N_{low}$. We propose a collision avoidance strategy in Protocol $1$, based on activity patterns and soft allocation. However, we do not assume sparsity as an activity pattern.} 
\color{black}
\begin{protocol}{MUSA Spreading Sequence Allocation}
\textit{Inputs:} Category of MTDs under two clusters (high-active $(N_{high})$ and low-active $(N_{low})$), $N$ of MUSA spreading sequences with known auto-correlation.\\
\textit{Goal:} Non-collision MUSA sequence allocation.\\
\textit{The protocol:}
\begin{enumerate}
  \item \textbf{Setup.}
  \begin{enumerate}
    \item
    Setup two clusters of MTDs $( N = N_{high} + N_{low}, \in \mathcal{N})$, define $\nu_{high} = \frac{N_{high}}{N},\nu_{low} = \frac{N_{low}}{N}$.

    \item
   Soft allocation of all available spreading sequences into two groups with the ratio:
    \begin{itemize}
        \item MUSA sequences for high active MTDs: $\frac{1}{\nu_{low}}$
        \item MUSA sequences for low active MTDs: $\frac{1}{\nu_{high}}$
    \end{itemize}
\textit{\textbf{soft allocation}}: set auto-correlation threshold for high and low active MTDs to choose the spreading sequences in the pre-defined separate group of MUSA spreading sequences.

\textit{\textbf{grouping}}: prioritize low-cross correlated MUSA spreading sequences during the grouping for high active MTDs.    
  \end{enumerate}
\end{enumerate}
\end{protocol}
\color{black}
\section{Deep Learning-Based MUD and Sparsity Estimation}
\subsection{Proposed Solution Approach}
This study aims to pave the way for the practical implementation of mMTC traffic MUD under the grant-free access scheme. Considering the sporadic nature of MTD activation pattern in mMTC traffic, only a small subgroup of the entire group of MTDs is active in a given time. By employing this fact, the MUD problem can be expressed as a deep learning-based multi-label classification. As the name suggests, multi-label classification detects the mutually non-exclusive multiple labels associated with the single data. In our problem, the labels are active and inactive states of the MTDs, and data is the received signal at the BS. 

The DNN-based MUD aims to detect the corresponding labels of the $\boldsymbol{\varphi}$. Here, the number of labels and their positions will be the number of active devices and their codebooks for SCMA and spreading sequences for MUSA. Therefore, we are not focusing on the recovery of the nonzero components. A major advantage of our proposed scheme is that since we train our network with sufficient training data, we do not need to estimate the channel before the detection. Therefore, we introduce a novel ResNet-based architecture to solve the tasks mentioned above jointly. Specifically, our architecture is not the same as conventional ResNet models, and we design layer-by-layer while minimizing the loss function and improving the probability of detection (recall). Here we improve the training using the novel residual unit shown in Fig.~\ref{fig:ResNet}. This architecture consists of several hidden nodes connecting the input layer to the output layer. Here each data item from the training data ($\mathbf{y}_{p}$) set is tagged with the corresponding label ($\boldsymbol{\varphi}$). To solve the two subtasks, the proposed model learns the mapping function associating the input nodes and related labels by updating the hidden parameters using the backpropagation process. Clearly, It is the nonlinear mapping $g$ between $\mathbf{y}_{p}$ and sparse vector elements of $\boldsymbol{\varphi}$. Subsequently, we reformulate the problem (\ref{argmin}) as
\begin{equation}
    \Omega = g(\mathbf{y}_{p};\omega,\theta),
    \label{reform}
\end{equation}
where $\omega$ is the set of weights of the hidden layers, and $\theta$ is the set of biases of the hidden layers of the neural network. The explicit intention of this DNN-MUD is to obtain $g$ characterized by $\omega$ and $\theta$ given $\mathbf{y}_{p}$, nearest to the unknown actual mapping function $g^*$. 
\subsection{MUD and Sparsity Estimation Architecture for SMV}
\begin{figure*}[tb]
\center
    \includegraphics[width=0.85\linewidth]{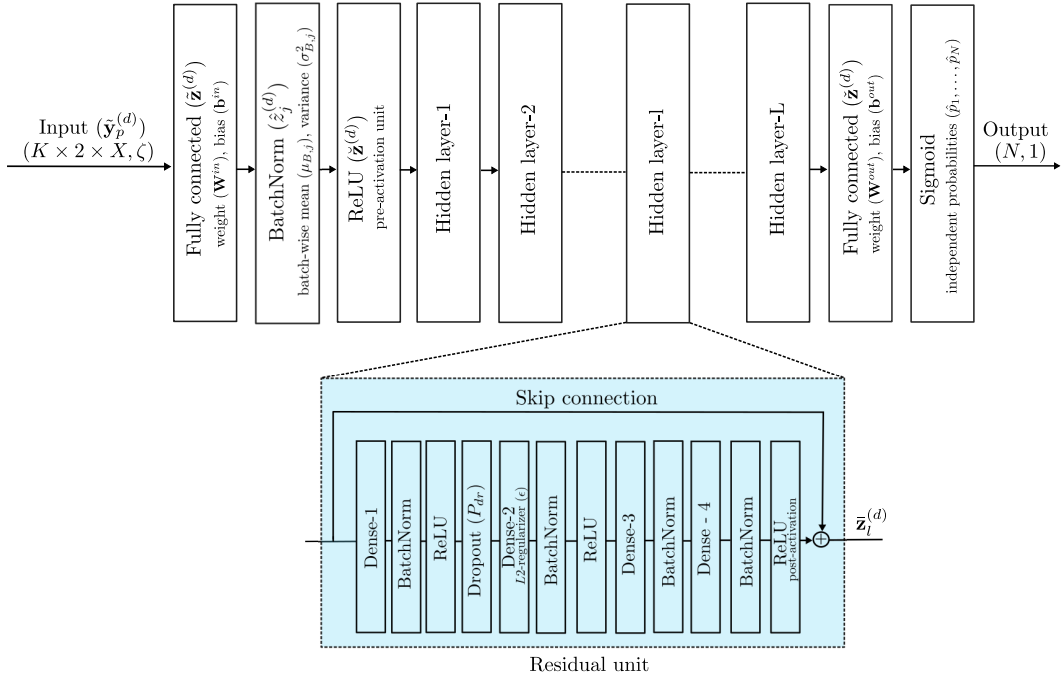}
    \caption{The complete configuration of the proposed ResNet concept-based MUD, including the layer structure of each hidden layer.}
    \label{fig:ResNet}
\end{figure*}

The proposed DNN-MUD architecture is designed based on the ResNet concept, which has several residual blocks. Each residual block consists of layers and functions such as dense, dropout, rectified linear unit (ReLU) activation unit, batch normalization, $L2$ regularization, and identity connection. DNN learns the parameters during backpropagation by calculating the error and gradient. Based on the gradient value, the weights of the hidden nodes are updated. This process continues until it reaches the input layer. The significant advantage of choosing ResNet over other architectures is to control the vanishing gradients where the identity connection holds it. During the training phase of the DNN-MUD, while learning the codebook parameters, it also learns the residual function. Clearly, DNN learns the difference between the input and the output, improving the system's accuracy until the residual value approaches zero. Therefore introducing the identity connection eases the training and improves the overall performance. Furthermore, the identity connection does not have any weight. Consequently, it does not include any parameters to the network; hence computational complexity remains the same as deep feedforward neural network. \textcolor{black}{In this study, we generate the set $\mathcal{D}$ of $D$ training data $(\Tilde{\mathbf{y}}_{p}^{(1)},\dots,\Tilde{\mathbf{y}}_{p}^{(D)})$ for each training iteration. Here, $\Tilde{\mathbf{y}}_{p}^{(d)}$ is a complex vector that cannot give direct input to the DNN because our label is not in the complex form. Therefore we split real and imaginary parts separately and stack them as a vector input to the system as
\begin{equation}
    \Tilde{\mathbf{y}}_{p}^{(d)} = [\Re(\Tilde{y}_{p,1}^{(d)}), \dots ,\Re(\Tilde{y}_{p,K}^{(d)}),\Im(\Tilde{y}_{p,1}^{(d)}), \dots ,\Im(\Tilde{y}_{p,K}^{(d)})]^T.
\end{equation}}
Fig.~\ref{fig:ResNet} shows the detailed structure of the proposed ResNet based DNN-MUD. First, the input data passes through the fully connected (FC) layer and the output vector \begin{math}\Tilde{\mathbf{z}}^{(d)} \in \mathbb{R}^{\upsilon \times 1}\end{math} of the FC layer is given by
\begin{equation}
    \Tilde{\mathbf{z}}^{(d)} = \mathbf{W}^{in}\Tilde{\mathbf{y}}_{p}^{(d)} + \mathbf{b}^{in}, \ \text{for} \ d = 1, \dots, D,
    \label{deep1}
\end{equation}
where \begin{math}\mathbf{W}^{in} \in \mathbb{R}^{\upsilon \times 2K} \end{math} is the initial weight, \begin{math}\mathbf{b}^{in} \in \mathbb{R}^{\upsilon \times 1}\end{math} is the initial bias and $\upsilon$ is the width of the hidden nodes. After that, $D$ number of resulting vectors are assembled in the mini-batch $\mathcal{B}$. Then, we add the batch normalization layer, where each element $\Tilde{z}_j^{(d)}$ in $\mathcal{B}$ is normalized to have zero-mean and unit-variance. The resulting vector $\Hat{\mathbf{z}}^{(d)}$ of the batch normalization layer~\cite{pmlr-v37-ioffe15} is stated as 
\begin{equation}
    \Hat{z}_j^{(d)} = \gamma\left(\frac{\Tilde{z}_j^{(d)}-\mu_{B,j}}{\sqrt{\sigma_{B,j}^2}}\right) + \eta,\ \text{for} \ j = 1, \dots, \upsilon,
        \label{deep2}
\end{equation}
where $\gamma$ and $\eta$ are scaling and shifting parameters, respectively. $\mu_{B,j}$ and $\sigma_{B,j}^2$ are the batch-wise mean and variance, respectively. The function of the normalization layer is to ensure that the input distribution has a fixed mean and variance, which improves the learning of the DNN. Also, larger variances of the training data reduce the learning of the internal features from the input signal. Therefore, batch normalization controls the interpretation affected by the various wireless channels and noise. Then, a nonlinear activation function is applied to the output of the batch normalization layer to determine whether the information passing through the hidden unit is activated. We added the ReLu activation~\cite{10.5555/3104322.3104425} unit before inputting the data to the residual units as the nonlinear activation function. The ReLU activation function is expressed by $\sigma(x) = max(x,0)$. Therefore the input to the first hidden layer is given by
\begin{equation}
    \Breve{\mathbf{z}}^{(d)} = \sigma\left(\Hat{\mathbf{z}}^{(d)}\right).
        \label{deep3}
\end{equation}
After the ReLU activation, the output vector passes through several identical blocks. Every block consists of four dense layers, four batch normalization layers, three ReLU activation units, a dropout layer, and a residual connection. Inside the identical block, the output of the first three layers follows as  (\ref{deep1}),  (\ref{deep2}),  (\ref{deep3}) with corresponding layer parameters. After that, a dropout layer is added to regularize the model to avoid the risk of overfitting and to distinguish the proper support. At the dropout layer, some activated hidden layer output units are randomly dropped out during the training by removing the connectivity of incoming and outgoing layers. The reason is the estimation of the  $\Omega$ depends on the activation patterns of the hidden units. Furthermore, if the sensing matrix $\boldsymbol{\Phi}$ is less correlated (low mutual coherence), then the detection of the active devices would be comparatively straightforward. However, when the number of active devices increases $\boldsymbol{\Phi}$ becomes highly correlated, which makes the estimation difficult. In effect, each update to a layer during training is performed with a different perspective of the configured layer. Therefore, the ambiguity of the activation patterns among correlated supports can be improved, which reduces the generalization error in the DNN. We define the dropout vector for the $l$-th residual unit as $\mathbf{q}^{(l)}$, then $j$-th component $q_j^{(l)}$ of the vector $\mathbf{q}^{(l)}$ can be expressed as
\begin{equation}
     q_j^{(l)} = \Bernoulli\left(P_{dr}\right),
\end{equation}
where $P_{dr}$ is the dropout probability of the Bernoulli random variable. In complex DNN models, there is a possibility that the model learns the noise in the training data and performs better during the training phase, but fails to detect during the testing phase where the noise is different. To avoid this, we added another dense layer with $L2$ regularizer after the dropout layer, where the regularizer introduces penalties to the specific layer parameters to avoid high variance. Here, we intend to introduce a decay function to the layer's weights, introducing a summed value to the regular loss function. Therefore, the updated loss function is given by
\begin{equation}
    \Tilde{\mathcal{L}}\left(\mathbf{W}^{l}_2\right) = \mathcal{L}\left(\mathbf{W}^{l}_2\right) + \frac{\epsilon}{2}\Vert \mathbf{W}^{l}_2\Vert^2_2,
\end{equation}
\textcolor{black}{where \begin{math} \mathcal{L}(.)\end{math} is the loss function,} \begin{math}\mathbf{W}^{l}_2 \in \mathbb{R}^{\upsilon \times \upsilon} \end{math} is the weights of the second dense layer of the $l$-th hidden block, and $\epsilon$ is the regularization ratio. During the training process, the $L2$ regularizer adds further subtraction on the current weights of the specific layer based on the $\epsilon$ value, which further improves the learning quality of the DNN. After that, we continue with the layers mentioned in Fig.~\ref{fig:ResNet}. Specifically, we include a ReLU activation at the residual unit's end to refine the path from one identical block to another. Therefore activated units are connected to improve the contribution of every identical block in the DNN-MUD. After having gone through all identical hidden layers, the last FC layer produces $N$ output values equal to the total number of MTDs. The output vector is given by
\begin{equation}
    \Check{\mathbf{z}}^{(d)} = \mathbf{W}^{out}\bigg(\Breve{\mathbf{z}}^{(d)} + \sum_{l=1}^L \Bar{\mathbf{z}}^{(d)}_l \bigg) + \mathbf{b}^{out},
\end{equation}
where $\Bar{\mathbf{z}}^{(d)}_l$ is the output of the hidden block $l$, \begin{math}\mathbf{W}^{out} \in \mathbb{R}^{N \times \upsilon} \end{math}, and \begin{math}\mathbf{b}^{out} \in \mathbb{R}^{N\times 1} \end{math} are the corresponding weight and bias of the last FC layer, respectively. After that, the sigmoid function produces $N$ independent probabilities $(\hat{p}_1, \dots , \hat{p}_N)$ based on the previous layer. The probability of the $j$-th MTD being active is given by
\begin{equation}
    \hat{p_j} = \frac{1}{1+e^{-z^{out}_j}},\ \text{for} \ j = 1, \dots, N.
    \label{deep4}
\end{equation}
Here, our motive is to convert the output probabilities into binary values, where each MTD's active state is indicated by one (and otherwise zero). A predefined probability threshold value exists \begin{math}(\geq 0.5) \end{math} in the sigmoid layer during the training phase. Therefore, we train our model to satisfy the threshold value by observing loss function and recall; hence, we do not need to define any hard threshold values to choose active devices in the testing phase. Final detection of active devices and sparsity is given by
\begin{equation}
    \Tilde{\Omega} = \arg\max_{\mid\Omega\mid=n} \sum_{j\in \Omega}\hat{p}_j.
    \label{DNN_final}
\end{equation}
Here, the value of $n$ and their positions solve the previously stated two subproblems. DNN-MUD learns the sparsity by properly annotating training labels during the training phase. Therefore, DNN-MUD satisfies (\ref{DNN_final}) and detects the active MTDs from the test data set. Consider $p_j$ represents the actual likelihood of the $j$-th MTD being operational. Hence, the cross entropy loss function of our model is given by
\begin{equation}
     \mathcal{L}({p_j},\hat{p_j}) = -\sum_{i = 1}^{N}{p_j}\log\hat{p_j}.
     \label{loss}
\end{equation}
\subsection{MUD and Sparsity Estimation Architecture for MMV}
Given that BSs are not limited by energy and computation complexity constraints, BSs in many practical network deployments are equipped with multiple antennas. We consider BS equipped with a \begin{math} X > 1 \end{math} antennas and MTDs equipped with a single antenna. Accordingly, BS receives multiple measurement vectors for each transmission. Therefore the previously mentioned SMV-based MUD becomes MMV-based MUD. Considering the facts, the received signal at the BS can be written as
 \begin{equation}
    \mathbf{Y}_{p} = \boldsymbol{\Phi}\boldsymbol{\Psi} + \mathbf{W}_p,
\end{equation}
where $\mathbf{Y}_{p} = [\mathbf{y}_{p,1} \dots \mathbf{y}_{p,X}]$ and $\boldsymbol{\Psi} = [\boldsymbol{\varphi}_1 \dots \boldsymbol{\varphi}_X]$. Here, the set of active MTDs at a given time slot is the same for all $X$ antennas. Therefore, the sparsity of $(\boldsymbol{\varphi}_1 \dots \boldsymbol{\varphi}_X)$ are the same. Hence, the MMV scenario gives additional information about the different channel parameters to support MUD. Therefore we are only considering changing the input data to our DNN-MUD while the labels remain unchanged. In this subsection, we generate the set $\Tilde{\mathcal{D}}$ of $D$ training data $(\Tilde{\mathbf{Y}}_{p}^{(1)},\dots,\Tilde{\mathbf{Y}}_{p}^{(D)})$ for each training iteration. Here, 
\textcolor{black}{$\Tilde{\mathbf{Y}}_{p}^{(d)}$ is a complex matrix that cannot input directly into the DNN because our label is not in the complex form. Therefore we split the matrix $\Tilde{\mathbf{Y}}_{p}^{(d)}$ into vectors $(\Tilde{\mathbf{y}}_{p,1}^{(d)} \dots \Tilde{\mathbf{y}}_{p,X}^{(d)})$ and split each of the vectors in the same way of SMV. We split real and imaginary parts of the vectors separately and stack them according to the antenna order, and create a vector input to the system as
\begin{equation}
\begin{split}
        \hat{\mathbf{y}}_{p}^{(d)} & = 
    [\{\Re(\Tilde{y}_{p,1,1}^{(d)}), \dots ,\Re(\Tilde{y}_{p,1,K}^{(d)}),\Im(\Tilde{y}_{p,1,1}^{(d)}), \dots ,\Im(\Tilde{y}_{p,1,K}^{(d)})\}, \\ & \dots,
    \{\Re(\Tilde{y}_{p,X,1}^{(d)}), \dots ,\Re(\Tilde{y}_{p,X,K}^{(d)}), \Im(\Tilde{y}_{p,X,1}^{(d)}), \dots ,\Im(\Tilde{y}_{p,X,K}^{(d)})\}]^T.
    \end{split}
\end{equation}}
After that, we use the same architecture as the SMV mentioned in Fig.~\ref{fig:ResNet} to support the MMV scenario with different layer parameters. In MMV, we have $X$-fold increase in the training data for the same label, which leads to improved MUD with the lower training overhead for the same performance level.

\section{Theoretical Analysis for Complexity and Convergence}
\subsection{Complexity Analysis}
We investigate the computational complexity of the proposed DNN-MUD to compare the complexity with the state-of-the-art and examine the feasibility of the algorithm's implementation. We calculate the floating-point operations (FLOPs), which include all floating-point operations of the FC layer, batch normalization layer, activation functions, dropout, and sigmoid function. First, FLOPs of the initial FC layer (\ref{deep1}), which include matrix multiplication of \begin{math}\mathbf{W}^{in} \in \mathbb{R}^{\upsilon \times 2K} \end{math} with \begin{math} \Tilde{\mathbf{y}}_{p}^{(d)} \in \mathbb{R}^{2K \times 1} \end{math} and a vector addition of \begin{math}\mathbf{b}^{in} \in \mathbb{R}^{\upsilon \times 1}\end{math} is given as
\begin{equation}
    \mathcal{C}_1 = (4K-1)\upsilon + \upsilon.
    \label{Comp1}
\end{equation}
Second, FLOPs of bath normalization layer (\ref{deep2}), which involves element-wise addition, element-wise multiplication, scaling, and shifting, is given by  
\begin{equation}
    \mathcal{C}_2 = 4\upsilon.
     \label{Comp2}
\end{equation}
Then, FLOPs of the pre-activation ReLU unit is stated as
\begin{equation}
    \mathcal{C}_3 = \upsilon.
     \label{Comp3}
\end{equation}
Next, there are $L$ hidden layers involved in the design. Each hidden layer contains four dense layers, four batch normalization layers, three ReLU activation units, a dropout layer (FLOPs = $\upsilon$), and a identity connection (FLOPs = $\upsilon$). Therefore, the total FLOPs of hidden layers are defined by 
\begin{equation}
    \mathcal{C}_4 = L\bigg( 4\big[ (2\upsilon-1)\upsilon + \upsilon \big] + 4(4\upsilon) + 3\upsilon + \upsilon + \upsilon \bigg).
     \label{Comp4}
\end{equation}
Also, the FLOPs of the final FC layer with \begin{math}\mathbf{W}^{out} \in \mathbb{R}^{N \times \upsilon} \end{math} and \begin{math}\mathbf{b}^{out} \in \mathbb{R}^{N\times 1} \end{math} is given as 
\begin{equation}
    \mathcal{C}_5 = (2\upsilon-1)N + N.
     \label{Comp5}
\end{equation}
Finally, the FLOPs of the sigmoid layer (\ref{deep4}), which contains four separate operations, is defined by
\begin{equation}
    \mathcal{C}_6 = 4N.
     \label{Comp6}
\end{equation}
Collectively, the computational complexity of the proposed DNN-MUD (From (\ref{Comp1}) to (\ref{Comp6})) in FLOPs is derived as 
\begin{equation}
\begin{split}
    \mathcal{C}_{DNN} & = 8L\upsilon^2 + (21L + 4K + 2N + 5)\upsilon + 4N,\\
      & = \mathcal{O}(L\upsilon^2).
     \end{split}
     \label{Comp}
\end{equation}
We further investigate the computational complexity of stOMP~\cite{6145475,6674179}, LS-BOMP~\cite{8968401}, C-AMP~\cite{8501573,6478821}, D-MUD~\cite{8968401}, and CNN-MUD~\cite{9462894} algorithms. The complexities in FLOPs are presented in Table~\ref{tabcomp}.
\begin{table*}[tb]
\caption{Computational Complexity}
\begin{center}
\begin{tabular}{lllllll}
\hline
Algorithm & DNN-MUD & stOMP~\cite{6145475,6674179} & LS-BOMP~\cite{8968401} & C-AMP~\cite{8501573,6478821} & D-MUD~\cite{8968401} & CNN-MUD~\cite{9462894}\\
\hline
Complexity & $\mathcal{O}(L\upsilon^2)$ & $\mathcal{O}(N \log N)$ & $\mathcal{O}(nK^2N)$ & $\mathcal{O}(NK\tau)$ & $\mathcal{O}(L\upsilon^2)$ & $\mathcal{O}(XN^2)$\\
\hline
\end{tabular}
\label{tabcomp}
\end{center}
\end{table*}
In addition, $\tau$ is the number of iterations in the C-AMP algorithm. We compare the computational complexities in Table~\ref{tabcomp}, it is clear that the complexity of the DNN-MUD and D-MUD depend on the internal parameters of the DNN while stOMP, LS-BOMP, C-AMP, and CNN-MUD rely on the system parameters. Therefore, it is obvious that using DNN algorithms in the mMTC systems is feasible and practical. Furthermore, the complexity of the DNN-MUD is significantly smaller than LS-BOMP and C-AMP algorithms.
\color{black}
\subsection{Labelsets selection and convergence}
Considering the massive number of total MTDs and a varying number of active devices, selecting a sufficient amount of data and labels are significant. In this subsection, we refer to the term labelset as a collection of states of each MTDs to the specific data sample. Here, each labelset is associated with one or more active devices. Generating data and labelsets for all combinations with sufficient channel realizations will be more computationally intensive. Conversely, randomly selecting a small subset may lead to increased false positives and false negatives during testing. In addition, choosing a minimum number of subsets which optimum for all combinations and performances is called as a set cover problem (SCP), which is NP-hard~\cite{Rokach}. Considering the associated information of the codebook/spreading sequence entries of the MTDs in the data, we explore the pair-wise correlation of the labelset to reduce the labelset selection (note that pair-wise correlation does not mean the correlated active MTDs). Lemma~\ref{lemma_1} shows sufficient labelsets to cover the entire labels. 
\begin{lemma}
The probability of all pairs in a label set \begin{math} L_\varphi = \{\lambda_1, \dots, \lambda_N \}\end{math} being included by $\alpha$ number of random labelsets is specified as: 
\label{lemma_1}
\end{lemma}
\begin{equation}
    P_s \leq 1 - 2\frac{\bigg((\delta - 1)S_1 - S_2\bigg)}{\delta(\delta - 1)},
\end{equation}
where
\begin{equation}
    S_1 = \Comb{N}{2}\Bigg(\frac{\Comb{N-2}{n} + 2\Comb{N-2}{n-1}}{\Comb{N}{n}}\Bigg)^\alpha,
\end{equation}
\begin{equation}
    S_2 = \frac{3\Comb{N}{3}}{(\Comb{N}{n})^\alpha}\Bigg(\frac{1}{4(N-1)}\bigg(\Comb{N-4}{n} + 4\Comb{N-4}{n-1} + 4\Comb{N-4}{n-2}\bigg)^\alpha + \bigg(\Comb{N-3}{n} + 3\Comb{N-3}{n-1} + \Comb{N-3}{n-2} \bigg)^\alpha \Bigg),
\end{equation}
\begin{equation}
    \delta = 2 + \left \lfloor \frac{2S_2}{S_1} \right \rfloor.
\end{equation}
\renewcommand\qedsymbol{$\blacksquare$}
\begin{proof}
The proof is given in Appendix A. 
\end{proof}

We use Lemma~\ref{lemma_1} to prove the theorem~\ref{theorem_1}, that the learning converge when $\alpha$ increases which ensure $P_D$ when $n$ increases. 
\begin{theorem}
The training of the reformulated problem in (\ref{reform}) converges when $\alpha$ increases. 
\label{theorem_1}
\end{theorem}
\renewcommand\qedsymbol{$\blacksquare$}
\begin{proof}
The proof is given in Appendix B. 
\end{proof}
\color{black}
\subsection{Training Data Generation and Implementation}
The proposed DNN-MUD is supervised learning, specifically, a multi-label classification problem that takes the labeled data as input and learns the mapping between the data and the corresponding labels. The well-trained and calibrated model can detect the new set of data labels. Therefore, a sufficient amount of training data is required for the convergence of the DNN model. We validate the adequate amount of data by the loss function of the DNN model. The proposed model can learn the codebook parameters from the actual received signal. However, we can train, validate, and test the DNN using synthetically generated data. The industrial application can directly feed the received signal with proper labels to the DNN model, which is feasible since this application is for uplink communication of mMTC.

In the generated data, the received signal contains the sparse input vector  $\boldsymbol{\varphi}$ and the sensing matrix $\boldsymbol{\Phi}$ where all environment channel properties and randomness are involved in the sparse vector, not in the sensing matrix. The core learning component of the DNN depends on the codebook for SCMA and spreading sequences for MUSA, which is known to the BS a priori. Hence, the synthetically generated data does not degrade the performance of the proposed DNN-MUD. Consequently, we generate the SCMA codebook and MUSA spreading sequences for the different simulation requirements. \textcolor{black}{To generate the channel for SMV and MMV systems, we consider the channel coefficient ($\text{g}_i$) between the $i$-th MTD and the BS as 
\begin{equation}
        \text{g}_i = \sqrt{\beta_i}h_{i,s},
\end{equation}
where $\beta_i$, $h_{i,s}$ characterize large-scale, small-scale fadings, respectively. Here, we assume \begin{math} h_{i,s}, i = 1, \dots, N \end{math}, are independent and identically distributed \begin{math} \mathcal{CN}(0,1)\end{math} random variable. Because the MTDs are distributed over a large area, the scatters around and between the BS and MTDs are different.} Then, create the channel vector and random noise vector $\mathbf{w}$. Finally, the training data is produced using (\ref{eq:generation}), and the sparse block vector's support is used as the training label. Furthermore, we train the DNN-MUD model offline for various environmental conditions with a different number of MTDs and a random number of active devices. Hence,  re-training the model is unnecessary unless new devices are added. Also, dropping out of devices does not influence the performance of our architecture.

\section{Simulation Results}
We explore the performances of the proposed DNN-MUD for SCMA and MUSA systems in this section. CSI is not assumed at the BS since the BS does not know the active users a priori and cannot extract the CSI from the pilot sequences. First, we show the results of the system performances of the DNN using a standard calibration curve to ensure correctness and reliability. Then, we present the outcomes for the SMV scenario in comparison with LS-BOMP~\cite{8570860}, C-AMP~\cite{6478821}, stOMP~\cite{6145475}, D-MUD~\cite{8968401}, and CNN-MUD~\cite{9462894}. The LS-BOMP minimizes the least-squares of a linear system with ~$\ell_0$ constraint and estimates the active devices with the knowledge of sparsity, C-AMP uses state evolution to estimate the active devices without the knowledge of sparsity, and stOMP evaluates the sparsity by sequentially transforming the received signal into a negligible residual in a fixed number of stages with the knowledge of the sparsity. After that, we demonstrate the plots of the MMV scenario. Finally, MUD results for the four different mmWave indoor factory environments are presented. The simulation parameters for the first four subsections are presented in Table~\ref{simulation1}. The number of neurons of the residual unit for SMV and MMV is presented in Table~\ref{neurons}
\begin{table*}[tb]
\caption{Simulation Parameters}
\begin{center}
\begin{tabular}{l l l l}
\hline
{Parameter}&  {Value}& {Parameter}&  {Value}\\
\hline
Overloading ratios (OR) & $150\%, 300\%$ &  Noise figure at BS & $4$ dB \\
Number of MTDs in SCMA ($N$) & $90$ & DNN sample size ($\zeta$) & $10^6$\\
Number of MTDs in MUSA ($N$) & $21$ & Number of hidden layers ($L$) & $4$ \\
Active probability ($p_i$) & $1/N$& Learning rate & $1\times 10^{-3}$\\
Log-normal shadow fading & $8$ dB & Batch size & $1 \times 10^{3}$\\
Noise spectral density& $-174$ dBm/Hz  & Epoch & $50$\\
 Transmission bandwidth & $1$ MHz & Drop out ($P_{dr}$) & $0.5$\\
 MTD to BS path loss model & $128.1 + 37.6\log_{10}(r_i[Km])$  & Regularization ratio ($\epsilon$) & $0.01$ \\
\hline
\end{tabular}
\label{simulation1}
\end{center}
\end{table*}

\begin{table*}[tb]
    \caption{Number of Neurons of the Residual Unit}
    \begin{center}
    \begin{tabular}{l l l l l l l l l l l l l}
       \hline  
       {} &          {SCMA} & {SCMA} & {MUSA} & {MUSA} & {SCMA} & {SCMA} & {MUSA} & {MUSA} & {SCMA} & {SCMA} & {MUSA} & {MUSA} \\
       {Layers} &    {150\%} & {300\%} & {150\%} & {300\%} & {150\%} & {300\%} & {150\%} & {300\%} & {150\%} & {300\%} & {150\%} & {300\%} \\
       {} &          {X = 1} & {X = 1} & {X = 1} & {X = 1} & {X = 2} & {X = 2} & {X = 2} & {X = 2} & {X = 4} & {X = 4} & {X = 4} & {X = 4} \\
       \hline
       Dense - 1 & {960} & {480} & {512} & {512} & {1920} & {960} & {512} & {512} & {3840} & {1920} & {512} & {512} \\
       Dense - 2 & {480} & {240} & {256} & {256} & {960} & {480} & {256} & {256} & {1920} & {960} & {256} & {256} \\
       Dense - 3 & {240} & {120} & {128} & {128} & {480} & {240} & {128} & {128} & {960} & {480} & {128} & {128} \\
       Dense - 4 & {120} & {60} &  {28}  & {14} & {240} &  {120} &  {56} &  {28} & {480} & {240} & {112} & {56} \\
       \hline  
    \end{tabular}
    \label{neurons}
    \end{center}
\end{table*}

 We use the binary cross-entropy as the loss function and the Adam optimizer to guarantee the model's convergence. The proposed DNN-MUD was designed, trained, validated, and tested on Keras with the Tensorflow backend. For the performance matrices, we use true positives $(\mathbf{tp})$, and true negatives $(\mathbf{tn})$, false positives $(\mathbf{fp})$, false negatives $(\mathbf{fn})$, and AUC to train and validate the proposed DNN-MUD model. Then, we examine the probability of detection (recall) on the testing set as $(P_D)$ = \big($\frac{\mathbf{tp}}{\mathbf{tp + fn}}$\big), probability of misdetection (false negative rate) as $(P_M)$ = \big($\frac{\mathbf{fn}}{\mathbf{tp + fn}}$\big), positive predictive values (precision) =  \big($\frac{\mathbf{tp}}{\mathbf{tp + fp}}$\big), and binary accuracy = \big($\frac{\mathbf{tp + tn}}{\mathbf{tp + fb + tn + fn}}$\big) to ensure the correctness of the probability of detection. 
 
\subsection{Performance of MUSA sequence selection algorithm and allocation protocol}
\begin{figure}
     \centering
     \begin{subfigure}[b]{0.46\linewidth}
         \centering
         \includegraphics[width=\linewidth]{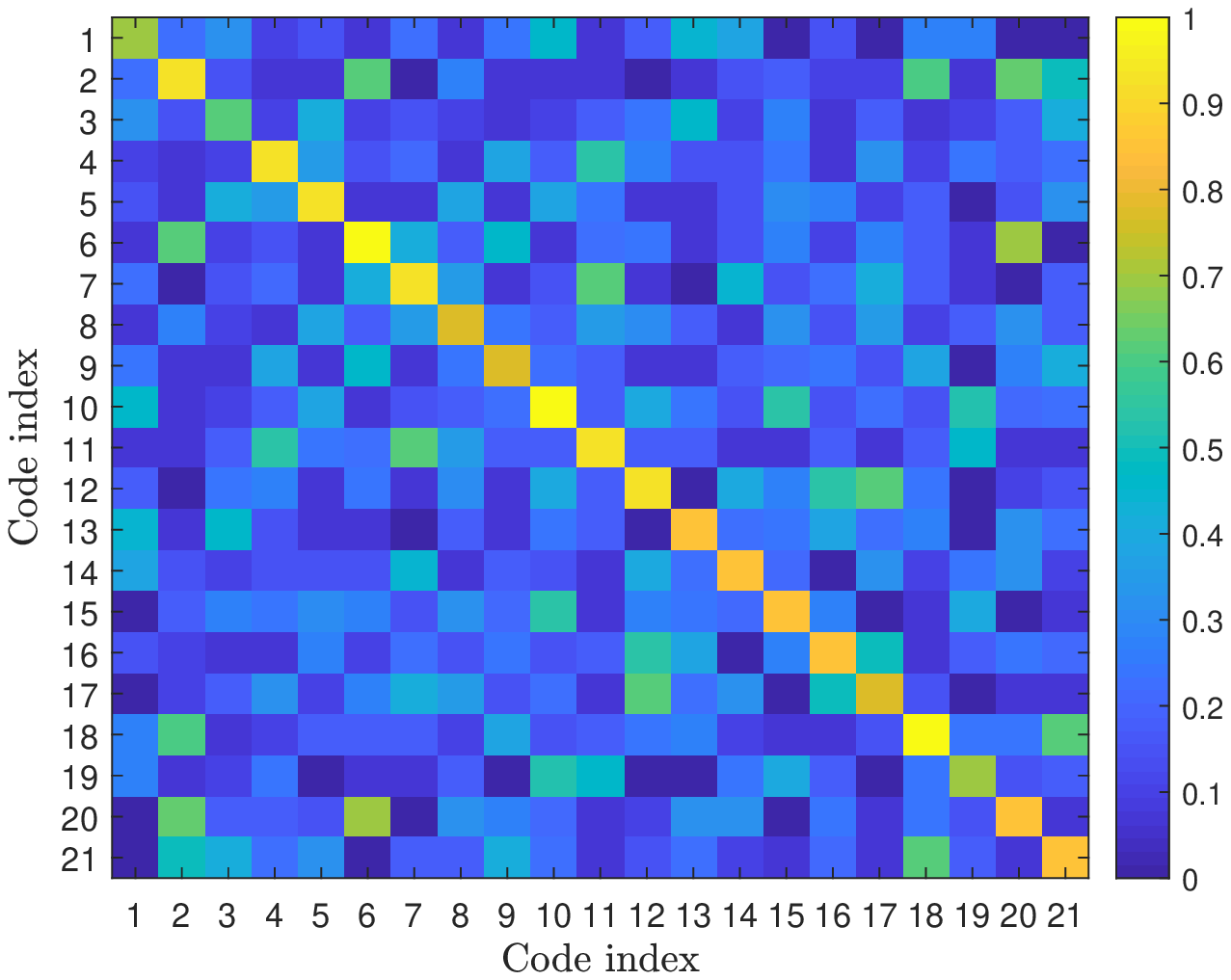}
         \caption{Proposed algorithm}
         \label{fig:Correlation_Algorithm}
     \end{subfigure}
     \hspace{-1em}
     \begin{subfigure}[b]{0.46\linewidth}
         \centering
         \includegraphics[width=\linewidth]{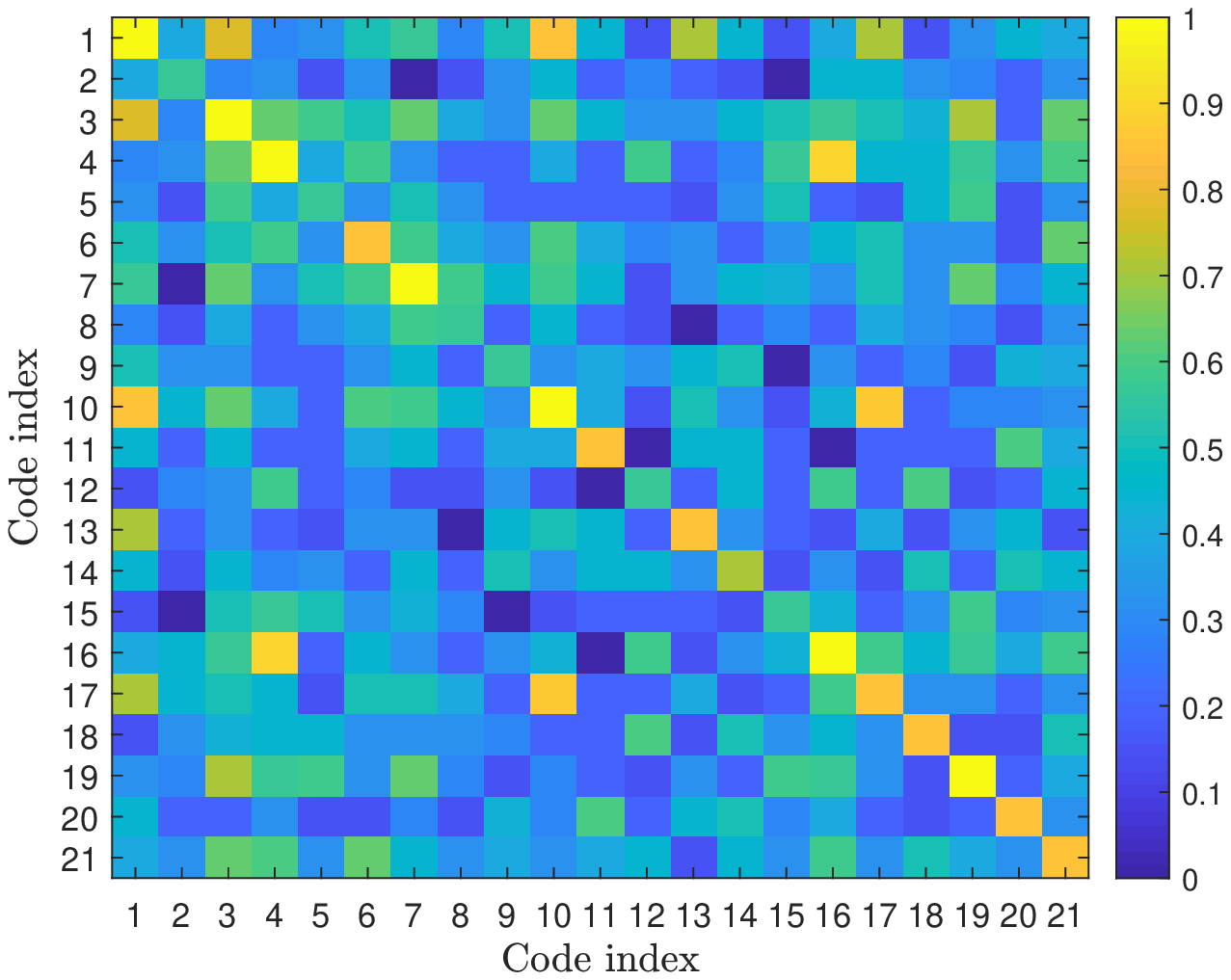}
         \caption{Random selection}
         \label{fig:Correlation_Random}
     \end{subfigure}
        \caption{Correlation of the MUSA sequences (a) selection by the proposed Algorithm 1 and (b) random selection.}
        \label{fig:Correlation}
\end{figure}

First, we consider the performance of the proposed MUSA sequence selection algorithm by plotting the correlation among the selected codewords. Fig.~\ref{fig:Correlation} compares the correlation of the MUSA sequences selected by the proposed Algorithm 1 (Fig.~\ref{fig:Correlation_Algorithm}) against a random selection of MUSA sequences from the available set (Fig.~\ref{fig:Correlation_Random}). The proposed algorithm shows a high degree of auto-correlation and low cross-correlation for the selected MUSA sequences as it chooses the sequences based on the orthogonality threshold value. In comparison, the randomly selected MUSA sequences do not exhibit a similar level of orthogonality. 
 \textcolor{black}{We evaluate the collision probability ($P_c$) for both random selection and the proposed protocol as below in Table~\ref{collision}. Example set up: $N = 100$, $N_{high} = 20$, $N_{low} = 20$, \begin{math} n \in N_{high} = 3\end{math}, and \begin{math} n \in N_{low} = 2\end{math}. Here, we can observe that the proposed scheme reduces the collision probability to a minimal value. Therefore, we assume no collision occurred during the initial access.}
\captionsetup[table]{labelfont=black,textfont=black}
\begin{table*}[h]
\color{black}
\caption{MUSA collision probability}
\begin{center}
\begin{tabular}{l l}
\hline
{Collision avoidance schemes}&  {$P_c$}\\
\hline
Random allocation & $5 \times 10^{-2}$ \\
Protocol $1$ & $3.125 \times 10^{-3}$ \\
\hline
\end{tabular}
\label{collision}
\end{center}
\end{table*}
\captionsetup[table]{labelfont=black,textfont=black}

\subsection{System Performance of DNN}
\begin{figure}
     \centering
     \begin{subfigure}[b]{0.46\linewidth}
         \centering
         \includegraphics[width=\linewidth]{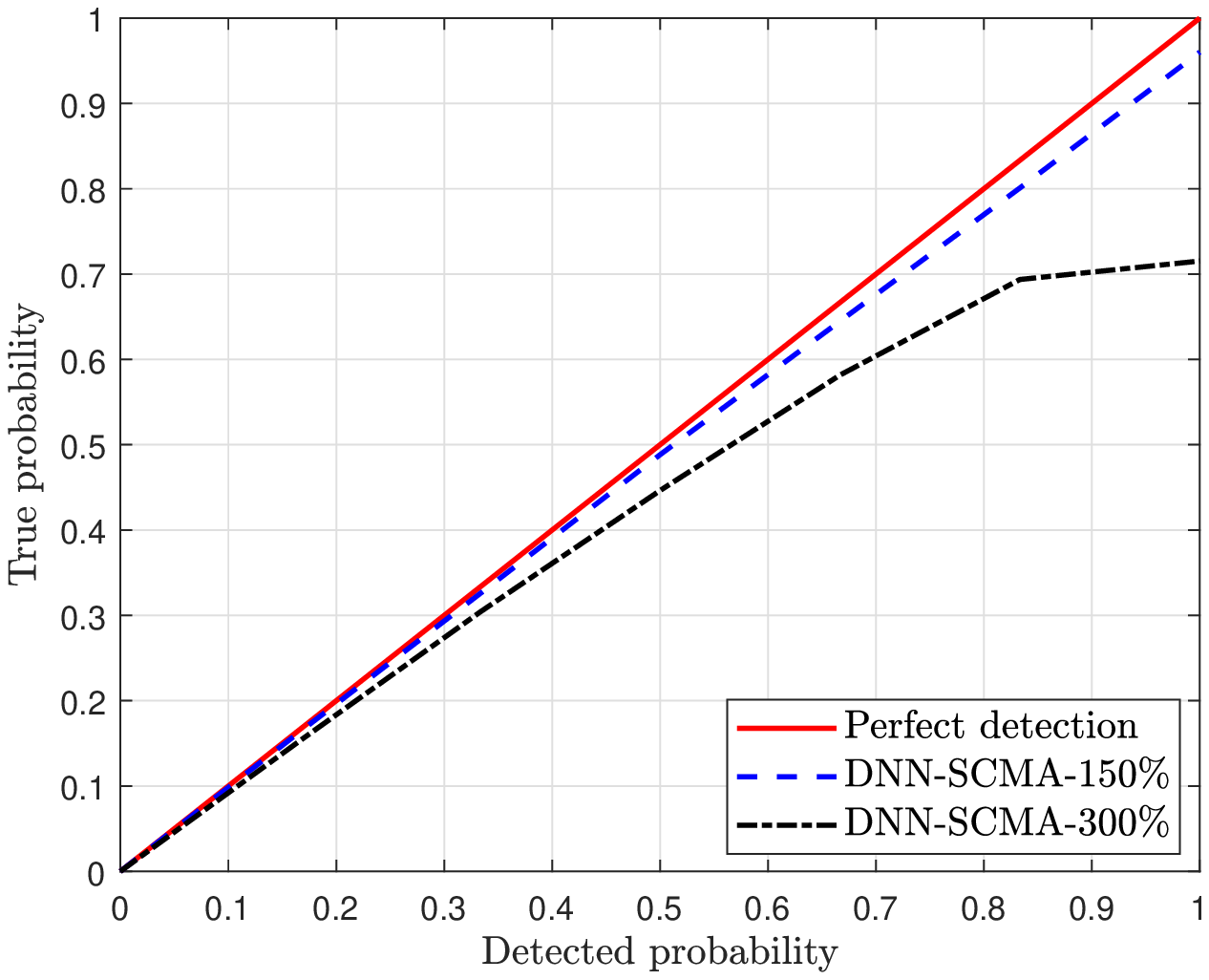}
         \caption{SCMA}
         \label{fig:Performance_SCMA}
     \end{subfigure}
     \hspace{-1em}
     \begin{subfigure}[b]{0.46\linewidth}
         \centering
         \includegraphics[width=\linewidth]{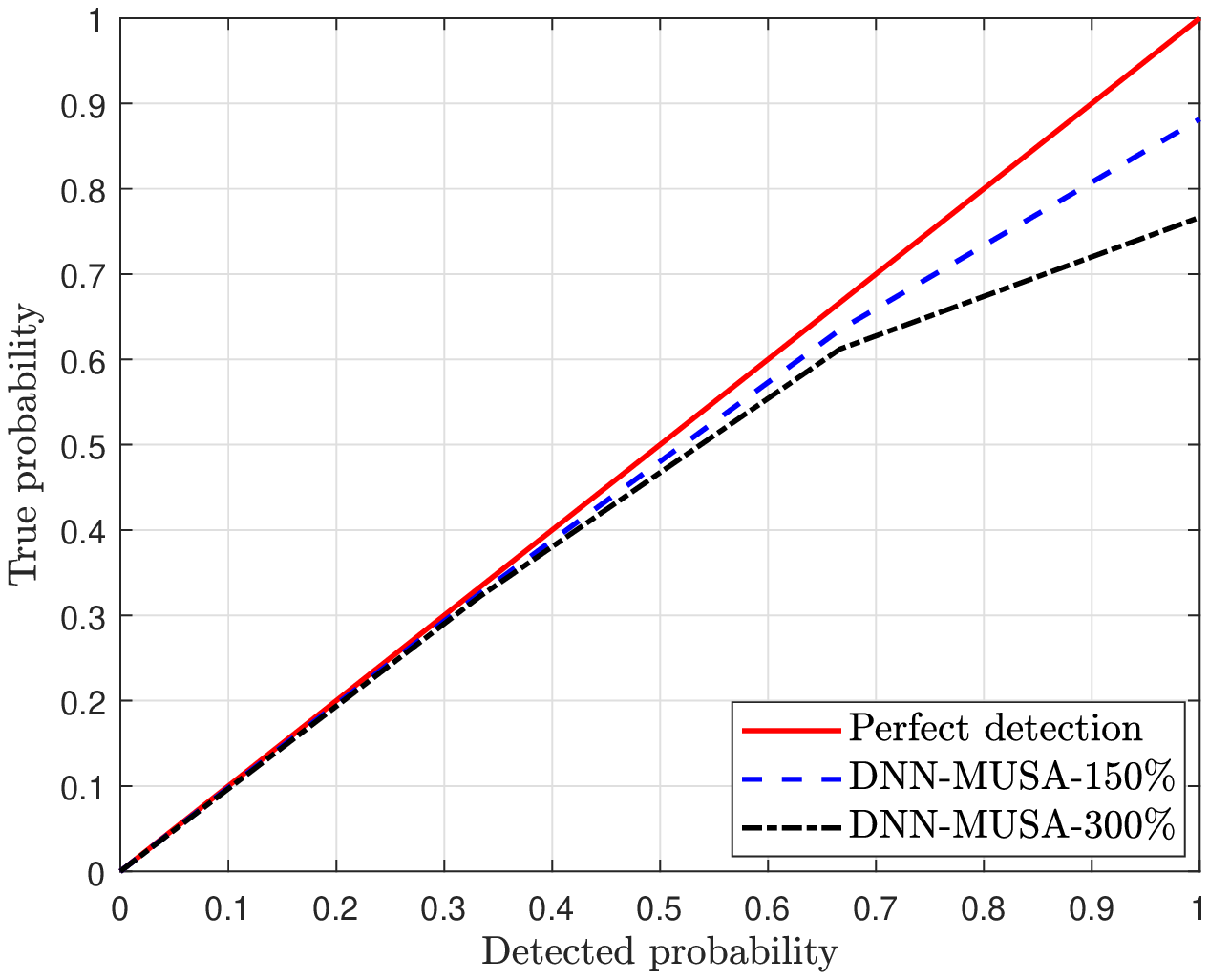}
         \caption{MUSA}
         \label{fig:Performance_MUSA}
     \end{subfigure}
        \caption{Calibration curve for DNN-MUD for both SCMA and MUSA systems with two different ORs (SNR = $20$\ dB, $X$ = $1$).}
        \label{fig:system_performance}
\end{figure}

We evaluate the performance of the DNN-MUD with the perfect detection of active devices. Fig.~\ref{fig:Performance_SCMA} and Fig.~\ref{fig:Performance_MUSA} show the calibration curve for the SCMA and MUSA systems, respectively. The calibration curve compares the expected output and the predicted output of the system. Furthermore, it is one of the standard approaches to evaluating system performance in multi-label classification. Our calibration curve clearly shows the perfectness of the DNN-MUD for OR = $150 \%$. In contrast, the requirement for the specific mechanism to improve the detection performance at higher ORs like $300 \%$ which we discussed in subsection D.
\subsection{MUD for the SMV Scenario}
\begin{figure}
     \centering
     \begin{subfigure}[b]{0.46\linewidth}
         \centering
         \includegraphics[width=\linewidth]{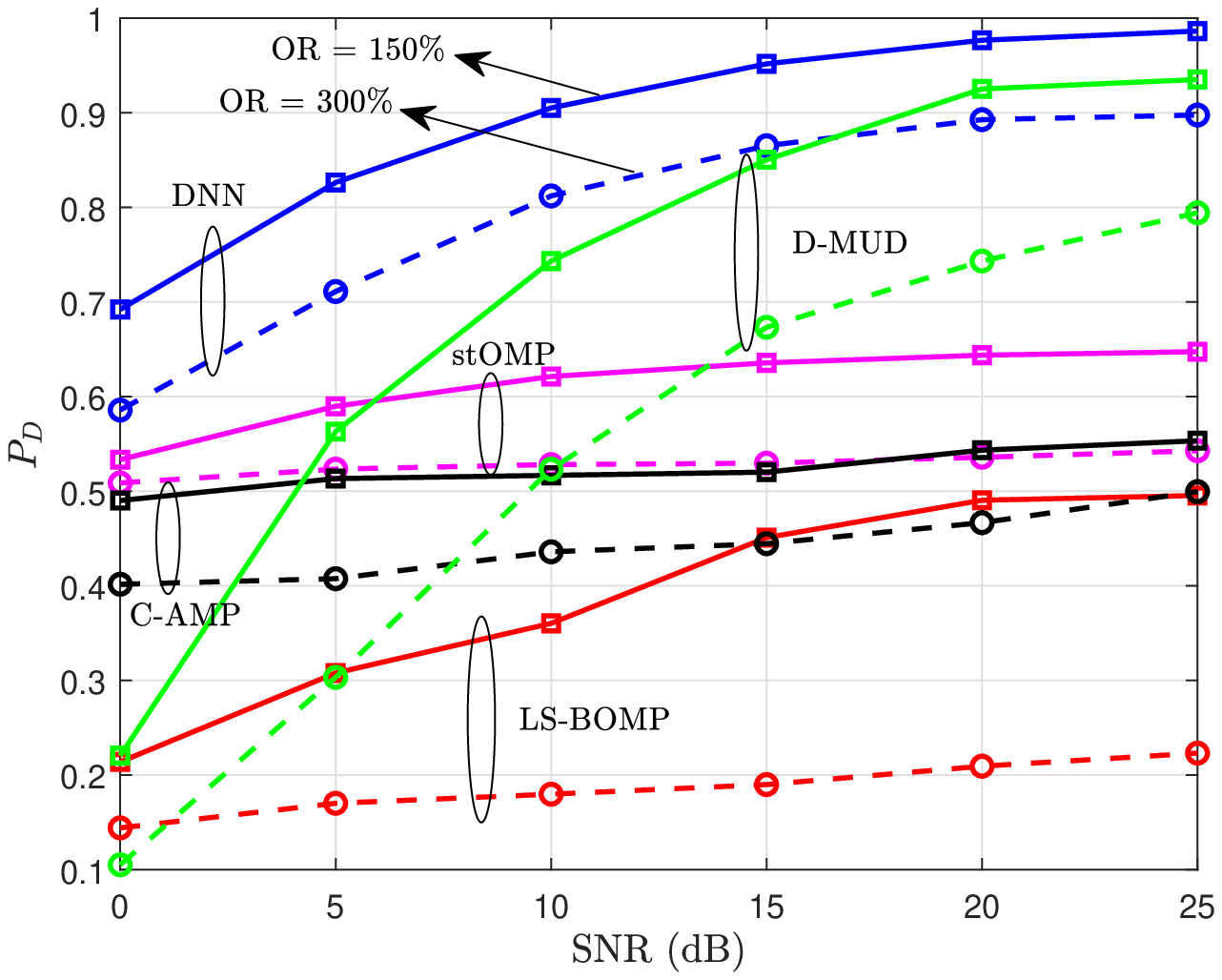}
         \caption{SCMA}
         \label{fig:SCMA_AD_3_SNR_All}
     \end{subfigure}
     \hspace{-1em}
     \begin{subfigure}[b]{0.46\linewidth}
         \centering
         \includegraphics[width=\linewidth]{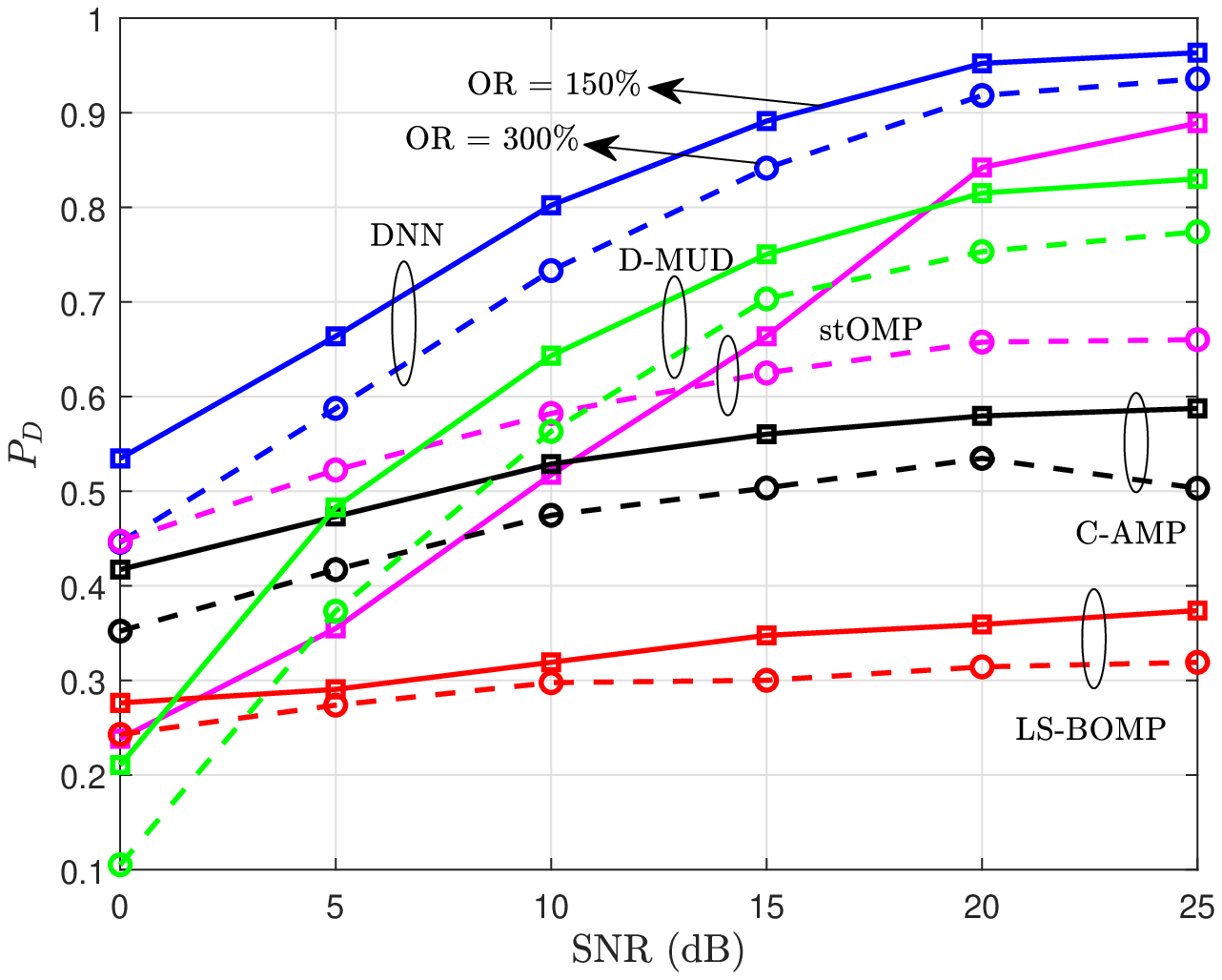}
         \caption{MUSA}
         \label{fig:MUSA_AD_2_SNR_All}
     \end{subfigure}
        \caption{Probability of detection (recall) versus the SNR for four different MUD schemes for (a) the SCMA codebook ($N$ = 90, $X$ = 1, $n$ = 3) and (b) the MUSA spreading sequences ($N$ = 21, $X$ = 1, $n$ = 2) with two different overloading ratios.}
        \label{fig:two_graphs1}
\end{figure}

First, we study the probability of detection of the proposed DNN for both SCMA and MUSA systems and show the result in Fig.~\ref{fig:two_graphs1}. We observe that DNN-MUD outperforms the stOMP, LS-BOMP, C-AMP, and D-MUD~\cite{8968401} algorithms for both SCMA and MUSA systems even though the OR increases from $150\%$ to $300\%$. In MUSA systems, stOMP is able to detect active devices to a certain extent with known sparsity, nevertheless unable to defeat the DNN-MUD. 

\begin{figure}
     \centering
     \begin{subfigure}[b]{0.33\linewidth}
         \centering
         \includegraphics[width=\linewidth]{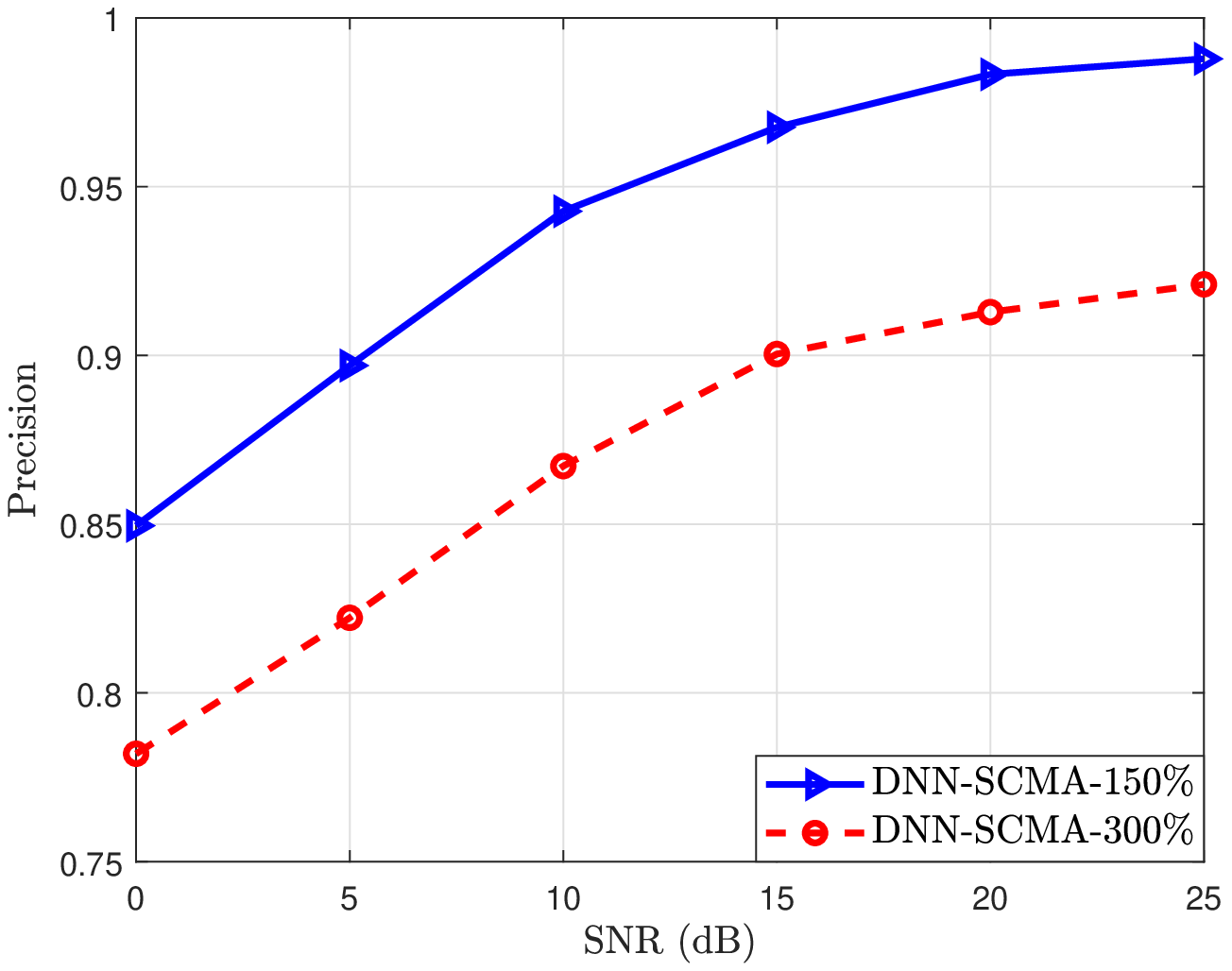}
         \caption{Precision}
         \label{fig:Precision_SCMA}
     \end{subfigure}
     \hspace{-1.5em}
     \begin{subfigure}[b]{0.33\linewidth}
         \centering
         \includegraphics[width=\linewidth]{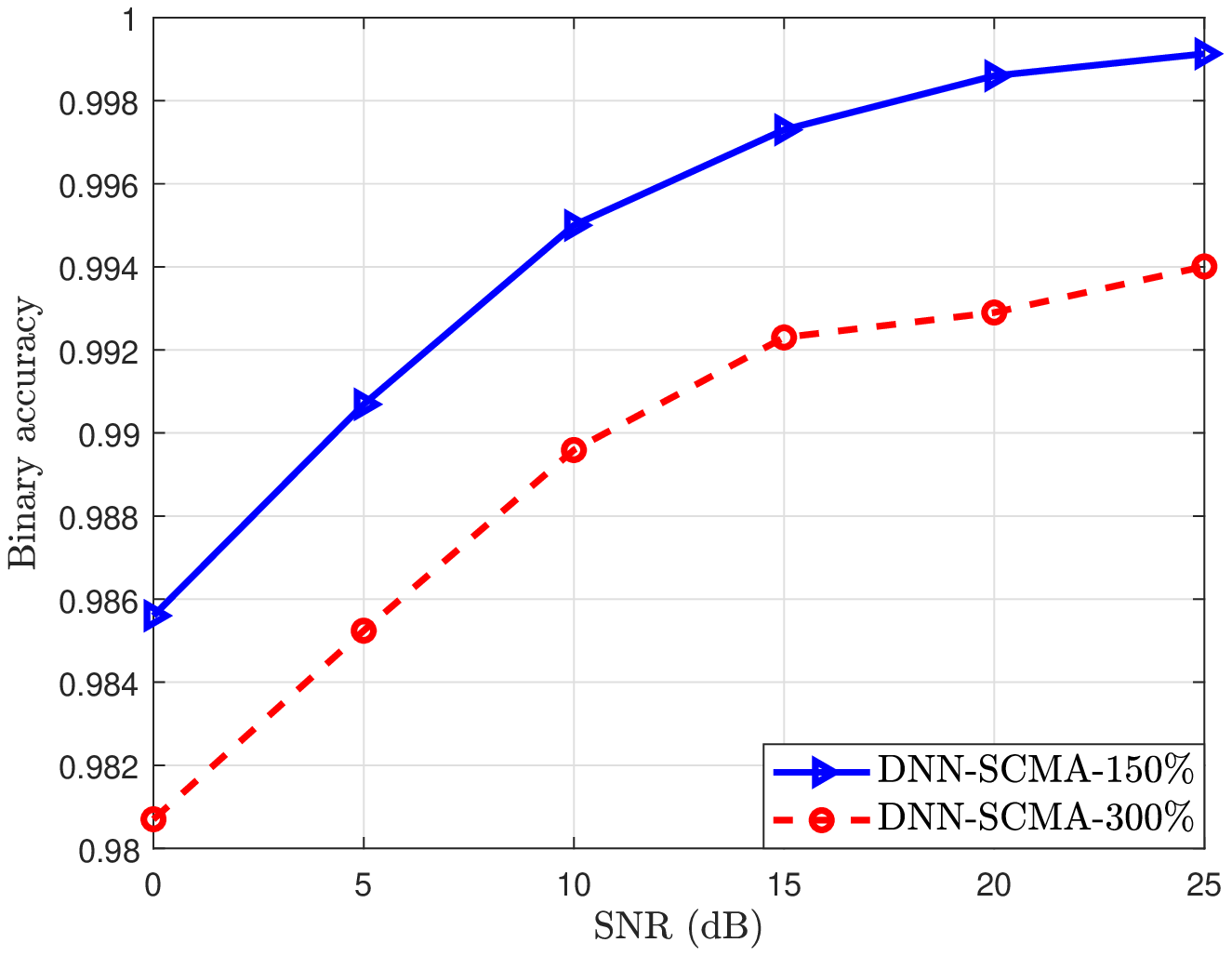}
         \caption{Accuracy}
         \label{fig:Accuracy_SCMA}
     \end{subfigure}
     \hspace{-1.5em}
     \begin{subfigure}[b]{0.33\linewidth}
         \centering
         \includegraphics[width=\linewidth]{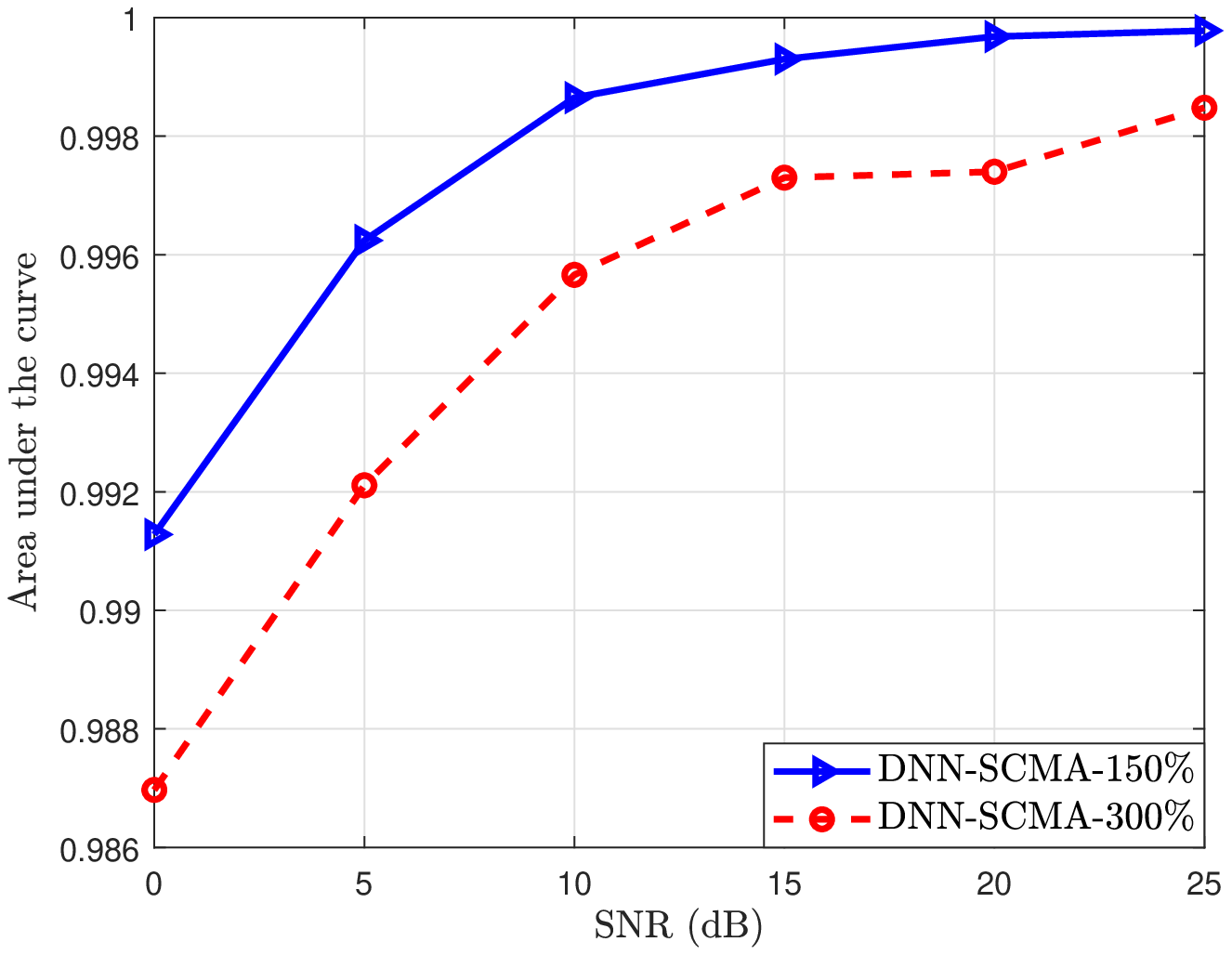}
         \caption{AUC}
         \label{fig:AUC_SCMA}
     \end{subfigure}
        \caption{DNN performance matrices for the SCMA scenario ($N$ = 90, $X$ = 1, $n$ = 3).}
        \label{fig:three graphs1}
\end{figure}

\begin{figure}
     \centering
     \begin{subfigure}[b]{0.33\linewidth}
         \centering
         \includegraphics[width=\linewidth]{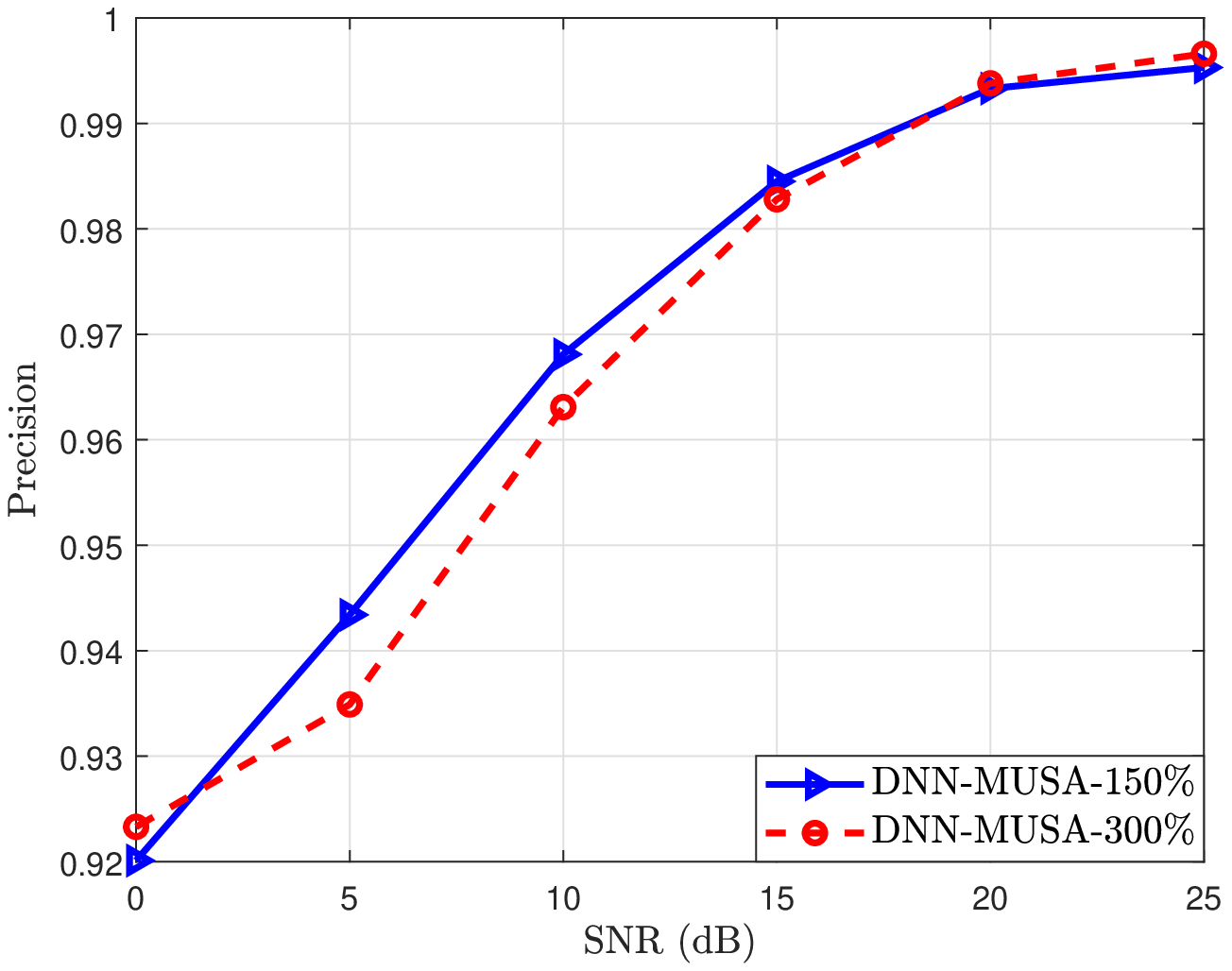}
         \caption{Precision}
         \label{fig:Precision_MUSA}
     \end{subfigure}
     \hspace{-1.5em}
     \begin{subfigure}[b]{0.33\linewidth}
         \centering
         \includegraphics[width=\linewidth]{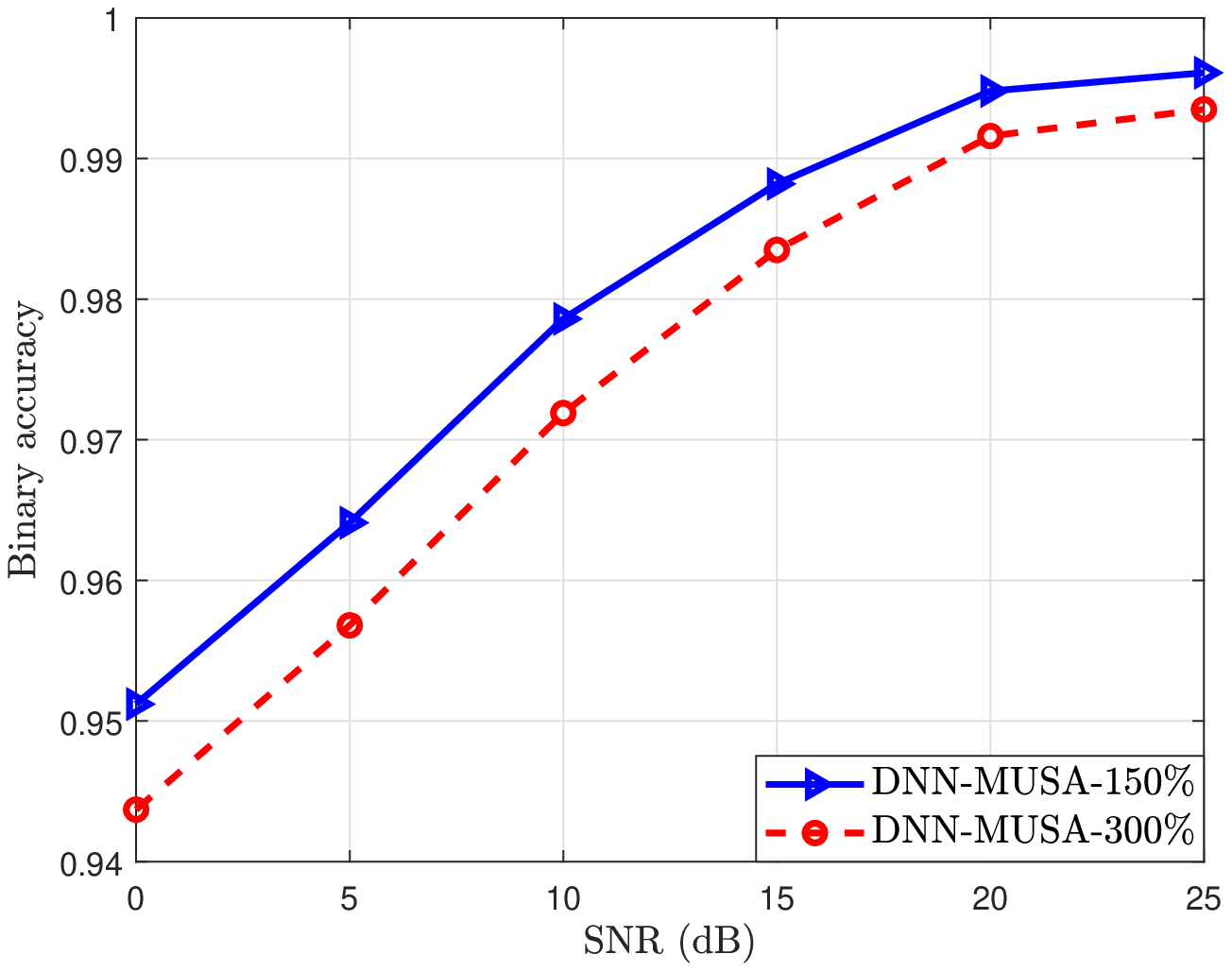}
         \caption{Accuracy}
         \label{fig:Accuracy_MUSA}
     \end{subfigure}
     \hspace{-1.5em}
     \begin{subfigure}[b]{0.33\linewidth}
         \centering
         \includegraphics[width=\linewidth]{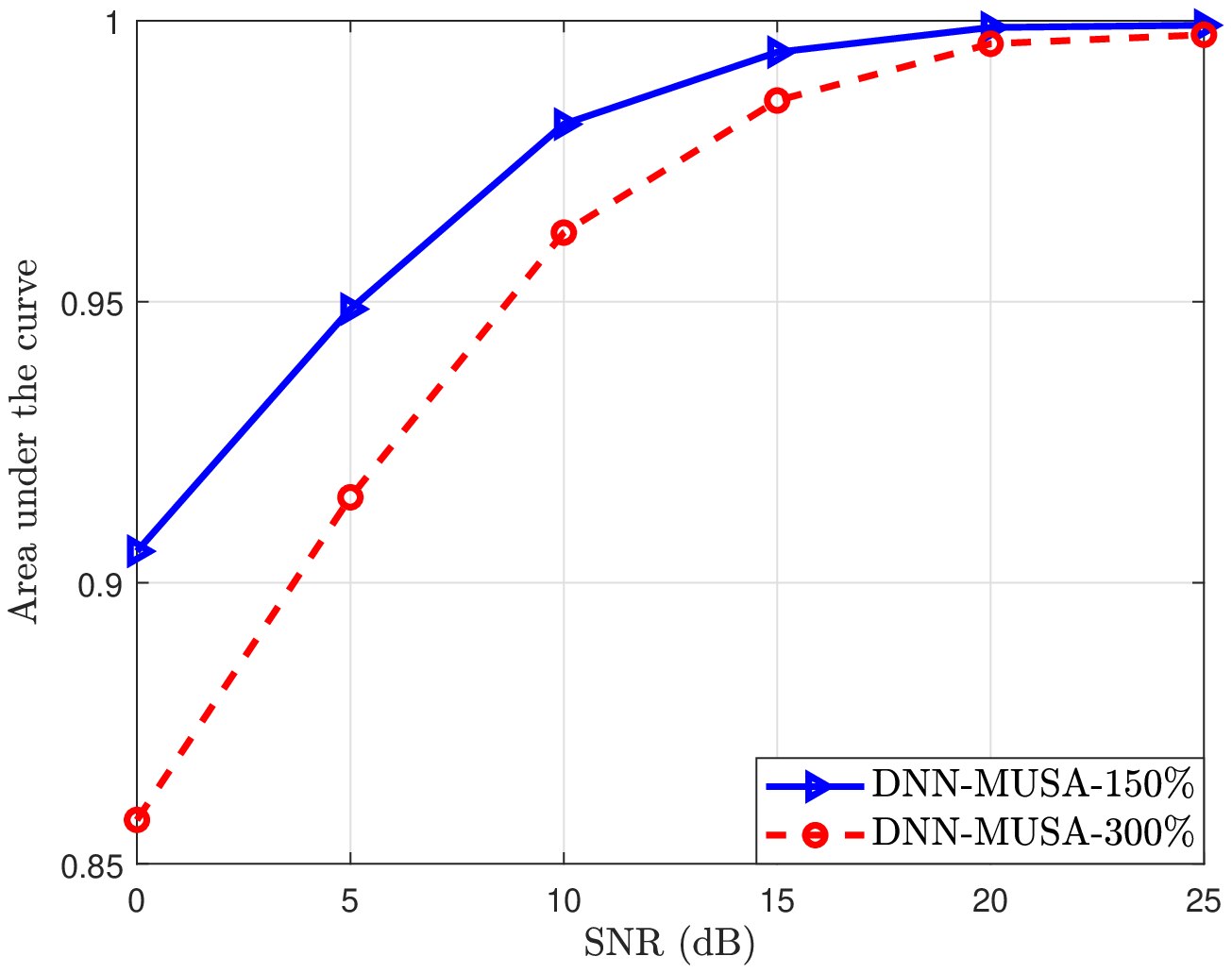}
         \caption{AUC}
         \label{fig:AUC_MUSA}
     \end{subfigure}
        \caption{DNN performance matrices for MUSA scenario ($N$ = 21, $X$ = 1, $n$ = 2).}
        \label{fig:three graphs2}
\end{figure}

We evaluate precision, binary accuracy, and AUC apart from the recall value to ensure the trustworthiness of the DNN and plot the results for the SCMA and MUSA system in Fig.~\ref{fig:three graphs1} and Fig.~\ref{fig:three graphs2}, respectively. Combining the outcomes of Fig.~\ref{fig:two_graphs1} with Fig.~\ref{fig:Precision_SCMA} and Fig.~\ref{fig:Precision_MUSA}, we understand the reliability of the DNN regarding the type-I (false positive) and type-II (false negative) errors. Furthermore, the binary accuracy of both SCMA and MUSA systems under two different ORs beyond $94\%$ indicates the overall performance. Likewise, the AUC reaches $1$ when SNR increases, which indicates the detection accuracy of our proposed approaches. 

\begin{figure}
     \centering
     \begin{subfigure}[b]{0.46\linewidth}
         \centering
         \includegraphics[width=\linewidth]{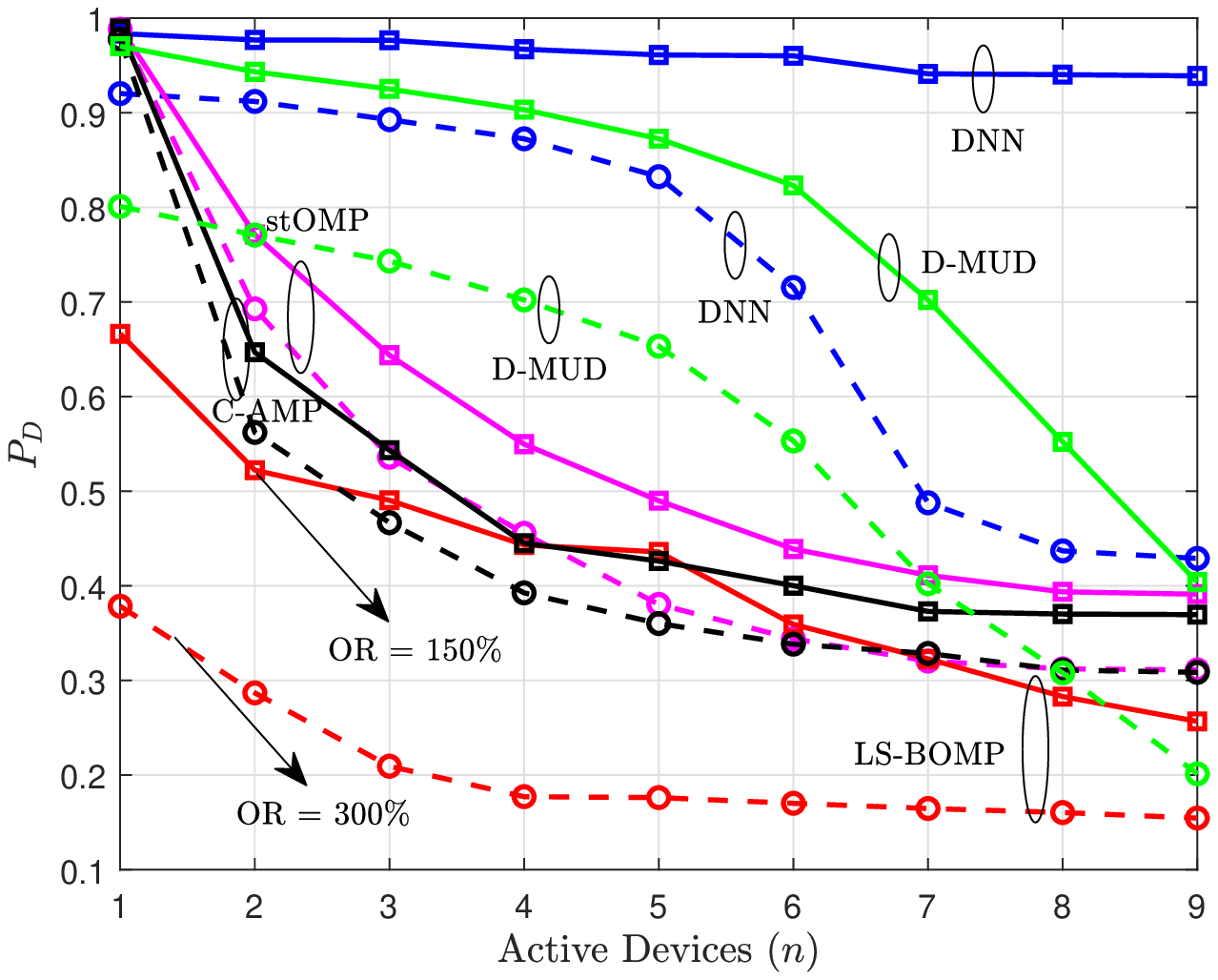}
         \caption{SCMA}
         \label{fig:Device_Activity_SCMA}
     \end{subfigure}
     \hspace{-1em}
     \begin{subfigure}[b]{0.46\linewidth}
         \centering
         \includegraphics[width=\linewidth]{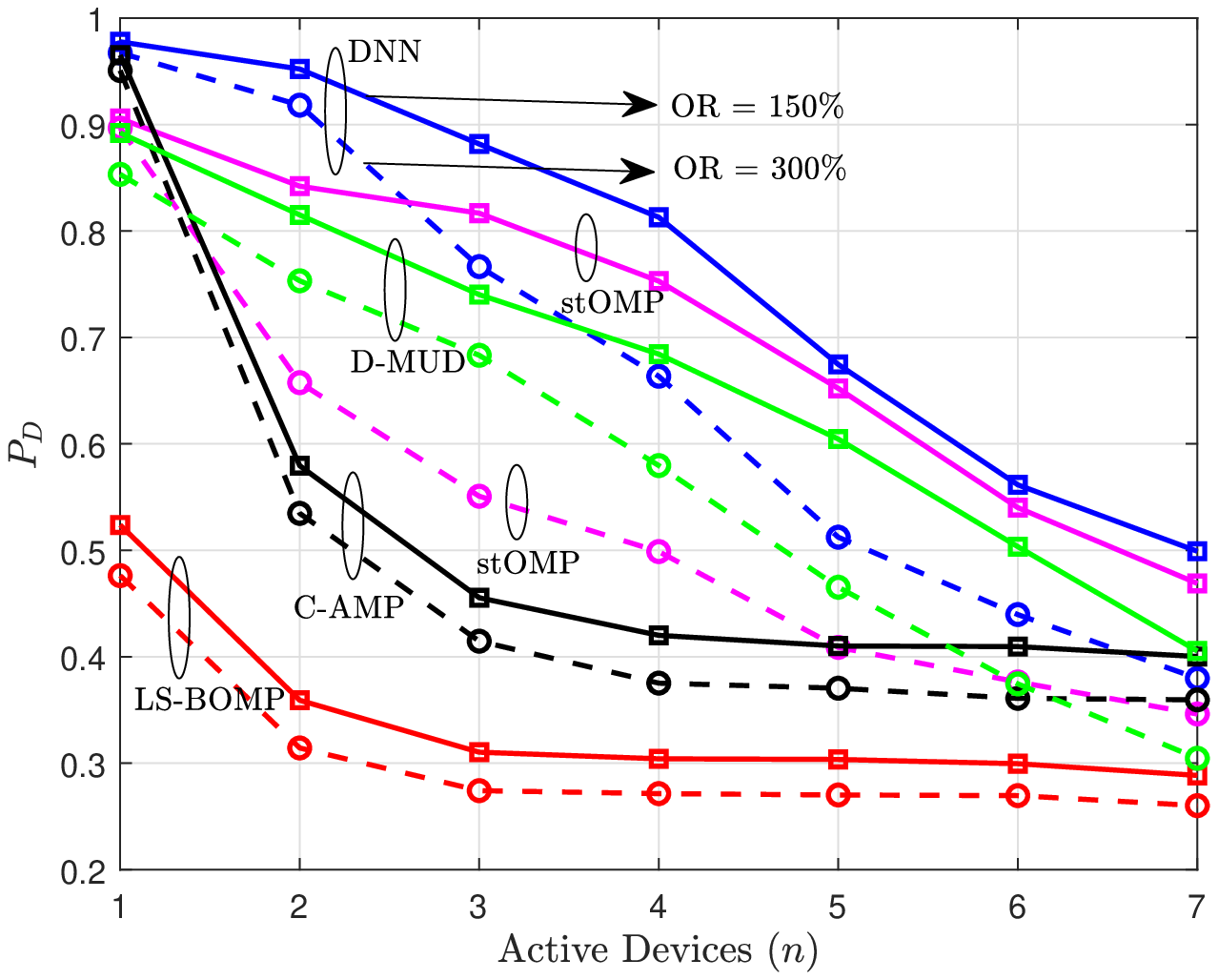}
         \caption{MUSA}
         \label{fig:Device_Activity_MUSA}
     \end{subfigure}
        \caption{Probability of detection (recall) versus the device activity for four different MUD schemes for (a) the SCMA codebook ($N$ = 90, $X$ = 1) and (b) the MUSA spreading sequences ($N$ = 21, $X$ = 1) with two different overloading ratios when SNR = $20$\ dB.}
        \label{fig:two_graphs2}
\end{figure}

Second, we explore the probability of detection versus the number of active MTDs at SNR = $20$\ dB and show the result in Fig.~\ref{fig:two_graphs2}. We observe that the DNN-MUD is superior to its counterparts in the entire region. Specifically, the examination was carried up to $10\%$ and $30\%$ of active devices out of all devices in SCMA and MUSA systems, respectively. Regardless, mMTC requirement of active devices is around $5 \%$ to $10 \%$ of total devices in a time frame. Furthermore, we can observe that the likelihood of detecting D-MUD~\cite{8968401} is around $87\%$, stOMP is around $50 \%$ and C-AMP is around $47 \%$ for OR = $150 \%$ in SCMA systems while DNN reaches more than $95 \%$. We can see the robustness of the DNN-MUD scheme with an increased number of active devices. According to the present results, when $n$ increases, the mutual coherence of the system rises sharply, leading to performance degradation. However, the DNN-MUD can handle this subject by learning the sensing matrix during the training phase, while stOMP, LS-BOMP, and C-AMP do not. 

\begin{figure}
     \centering
     \begin{subfigure}[b]{0.46\linewidth}
         \centering
         \includegraphics[width=\linewidth]{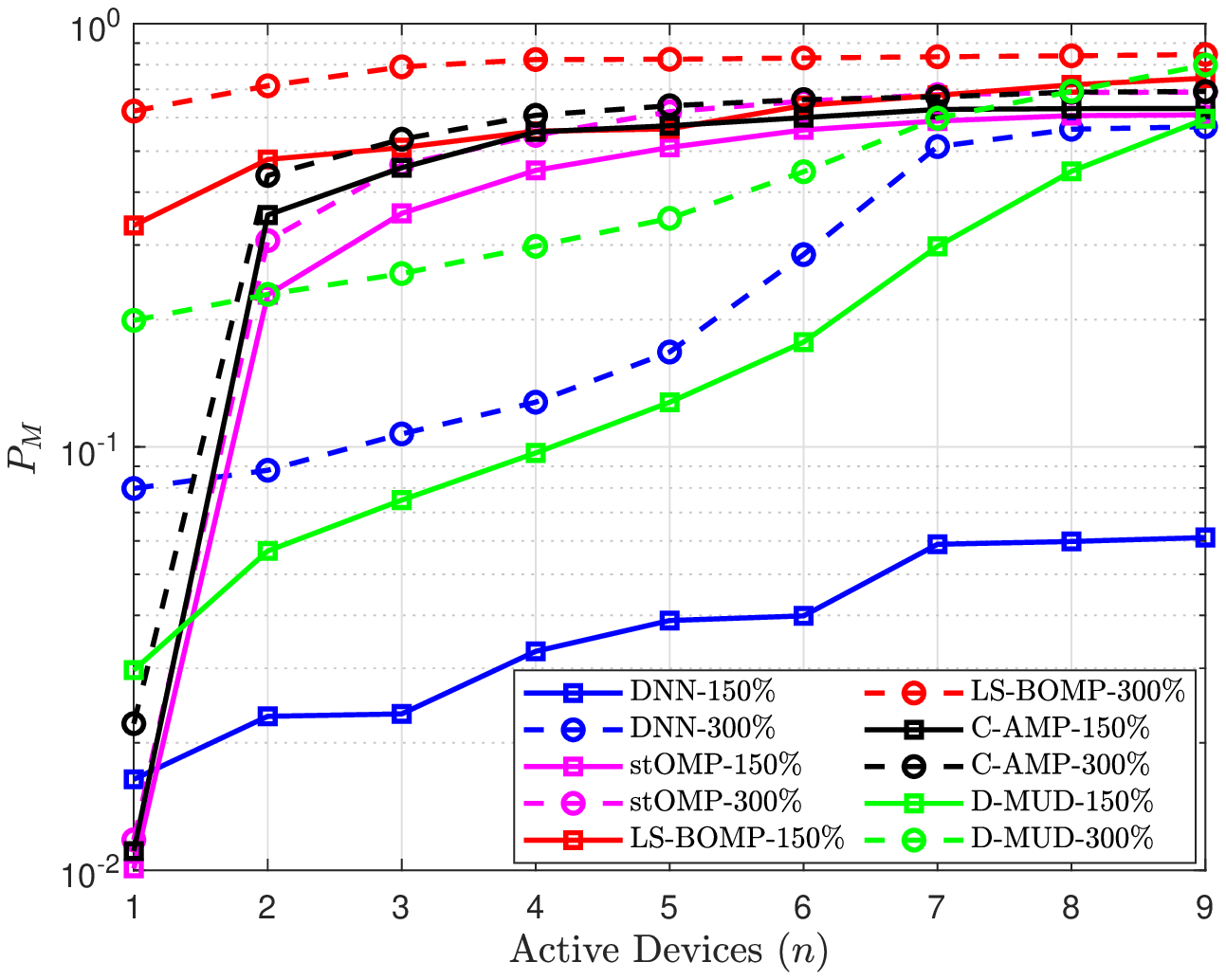}
         \caption{SCMA}
         \label{fig:Misdetection_SCMA}
     \end{subfigure}
     \hspace{-1em}
     \begin{subfigure}[b]{0.46\linewidth}
         \centering
         \includegraphics[width=\linewidth]{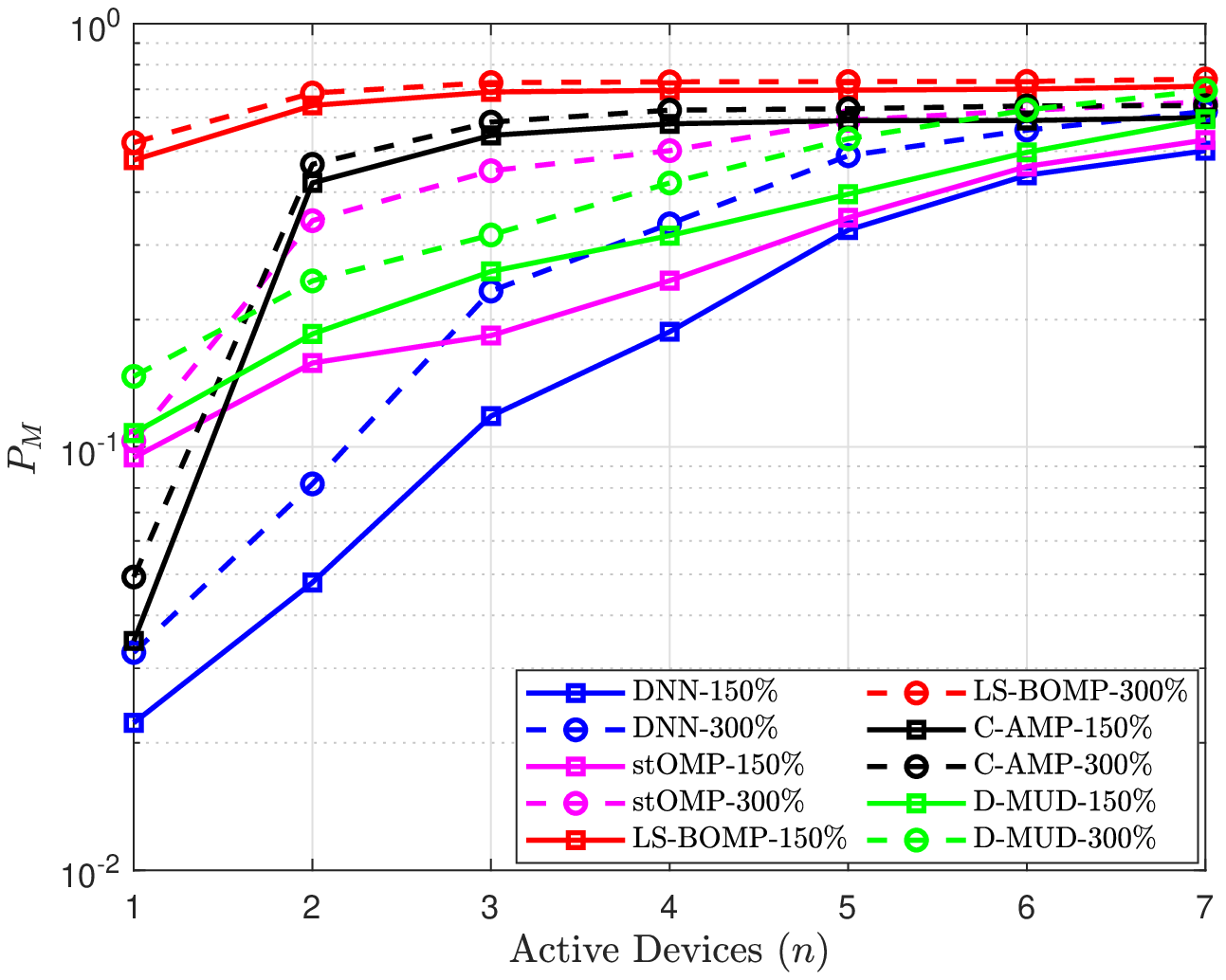}
         \caption{MUSA}
         \label{fig:Misdetection_MUSA}
     \end{subfigure}
        \caption{Probability of misdetection versus the device activity for four different MUD schemes for (a) the SCMA codebook ($N$ = 90, $X$ = 1) and (b) the MUSA spreading sequences ($N$ = 21, $X$ = 1) with two different overloading ratios when SNR = $20$\ dB.}
        \label{fig:two_graphs3}
\end{figure}

Finally, we investigate the probability of misdetection (false-negative rate) versus the number of active devices at SNR = $20$\ dB for DNN-MUD compared to stOMP, LS-BOMP, C-AMP and D-MUD~\cite{8968401} algorithms which are presented in  Fig.~\ref{fig:two_graphs3}. This outcome clearly explains that even though the number of active devices increases, the false-negative rate does not defeat by other algorithms for SCMA and MUSA systems under different OR scenarios. For example, there are $164$ misdetections out of $10^4$ samples when $n = 1$ while there are $611$ misdetections out of $10^4$ samples when $n = 9$ for SCMA codebooks with $150 \%$ ORs. 
\subsection{MUD for the MMV Scenario}
\begin{figure}
     \centering
     \begin{subfigure}[b]{0.46\linewidth}
         \centering
         \includegraphics[width=\linewidth]{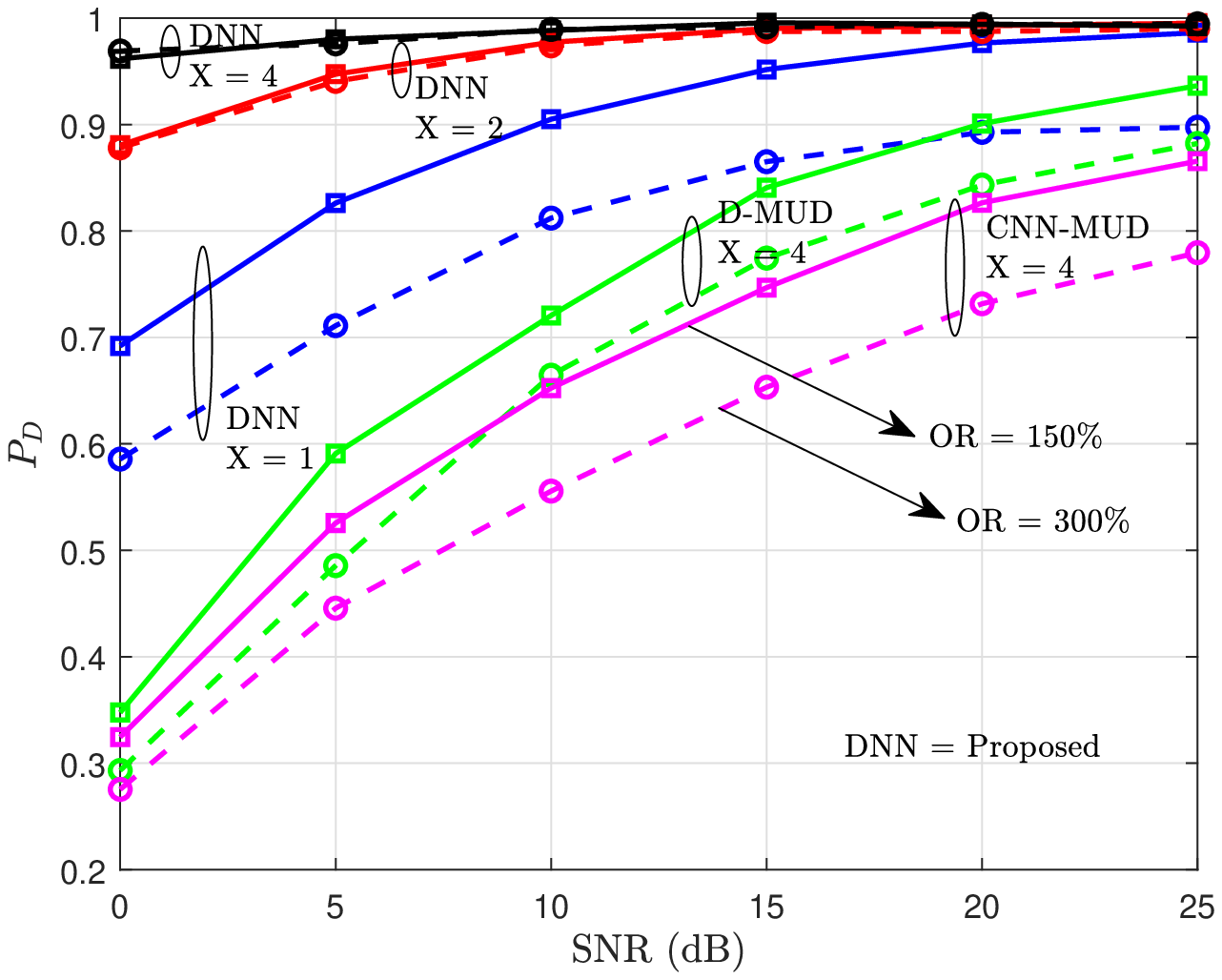}
         \caption{SCMA}
         \label{fig:MMV_SCMA}
     \end{subfigure}
     \hspace{-1em}
     \begin{subfigure}[b]{0.46\linewidth}
         \centering
         \includegraphics[width=\linewidth]{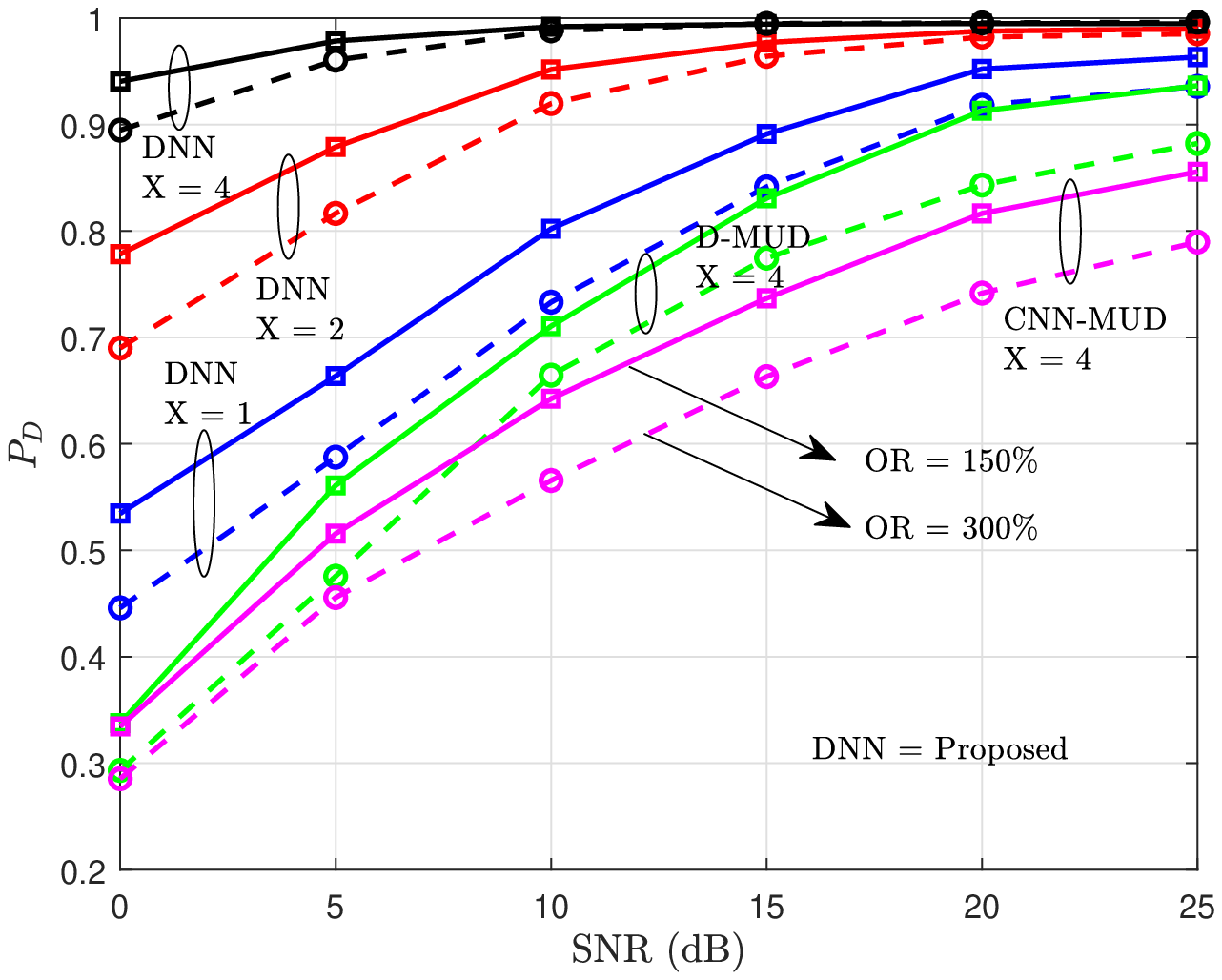}
         \caption{MUSA}
         \label{fig:MMV_MUSA}
     \end{subfigure}
        \caption{Probability of detection (recall) versus the SNR for two different overloading ratios for (a) SCMA codebooks ($N$ = 90, $n$ = 3) and (b) MUSA spreading sequences ($N$ = 21, $n$ = 2) with three different number of antenna configurations.}
        \label{fig:two_graphs4}
\end{figure}

Next, we evaluate the performance of DNN-MUD for MMV systems. Fig.~\ref{fig:two_graphs4} presents the variation of the probability of detection versus SNR with three different antenna configurations for both SCMA and MUSA systems; when $X$ increases, the spatial diversity increases, improving the detection probability. For example, at SNR = $10$\ dB the probability of detection improves from $81 \%$ to $97 \%$ when $X$ changes from $X = 1$ to $X = 2$ for SCMA codebooks with $300 \%$ OR. Likewise, for MUSA spreading sequences with $300 \%$ OR, the performance improves from $73 \%$ to $92 \%$. Moreover, the probability of detection reaches $100 \%$ for both SCMA and MUSA systems with $150 \%$ and $300 \%$ ORs, when $X$ goes to $X = 4$ at SNR = $15$\ dB. In addition, we compare the performances with D-MUD~\cite{8968401} and CNN-MUD~\cite{9462894} when $X = 4$, where our proposed DNN-MUD performs more promising than existing deep learning solutions. Here, we design our DNN such that with twice the number of input neurons when $X$ increases from one to two and two to four to achieve the advantage of the spatial diversity of the MMV system. Furthermore, the increased input neurons improve the learning capability of the DNN. Notably, the MMV system is the proactive solution to the issue we observed and discussed in Fig.~\ref{fig:system_performance} for higher OR ratios. Based on the outcomes, we can state that increasing the number of antennas at the BS is ideal for overcoming the drawbacks of higher ORs of SCMA and MUSA systems, which are applicable in mMTC for future wireless communication requirements. 
\subsection{MUD for the mmWave Indoor Factory Environment}
In this subsection, we consider the next-generation mMTC application environment. 5G new radio spectrum includes the 26 GHz mmWave band, while IEEE 802.11 ad/ay is defined for the 60 GHz band (known as WiGig or 60 GHz Wi-Fi). Future beyond 5G/6G wireless networks are expected to operate at even higher frequency bands~\cite{BroadbandWhitePaper2020}. Thus, there is a requirement to investigate high-frequency channel models. 
In that respect, we propose the SMV DNN-MUD problem for mmWave channel models based on the 3GPP specifications. We consider the indoor factory (InF) environment with various sizes and various density levels of machinery. 3GPP proposes five types of channel models for InF, first is sparse clutter with low BS height (InF-SL), second is dense clutter with low BS height (InF-DL), third is sparse clutter with high BS height (InF-SH), fourth is dense clutter with high BS hight (InF-DH), and the fifth is for 100\% line of sight (LOS) scenario~\cite{3gpp.138.901}. We choose the first four categories for further study. Furthermore, the sparse clutter environment consists of large metallic machinery placed inside the factory with sufficient clearance between the big machines with a considerable amount of open space available. Meanwhile, the dense clutter scenario contains small and medium size of metallic irregular types of machinery. In all four InF scenarios, we consider the LOS and non-LOS (NLOS) based on the LOS probabilities defined by 3GPP~\cite{3gpp.138.901}. In accordance, the $3D$ distance $r_{3D}$ between BS and the MTD is given as
\begin{equation}
    r_{3D} = \sqrt{r_{2D}^2 + {(h_{BS}-h_{MTD})}^2},
\end{equation}
where $r_{2D}$ is the $2D$ distance between the BS and the MTD, $h_{BS}$ is the height of the indoor BS, and $h_{MTD}$ is the height of the location where the MTD is deployed. The LOS pathloss of the all four InF scenarios define as 
\begin{equation}
    PL_{LOS} = 31.84 + 21.50\log_{10}\left(r_{3D}\right)+ 19.00\log_{10}(f_c) + \chi_{\sigma_{LOS}},
\end{equation}
where $f_c$ is the normalized carrier frequency and with the corresponding shadow fading standard deviation $\sigma_{LOS} = 4.3$. The NLOS pathloss define as
\begin{equation}
    PL = a + b\log_{10}\left(r_{3D}\right)+ 20.00\log_{10}(f_c) + \chi_{\sigma_{NLOS}},
\end{equation}
\begin{equation}
    PL_{NLOS} = \max(PL_{LOS},PL),
\end{equation}
where values of $a$ and $b$ varies with different InF scenarios. The values of $a$, $b$, and shadow fading standard deviations $\sigma_{NLOS}$ are given in Table~\ref{tab2}. The LOS probabilities of all four InF scenarios define as
\begin{equation}
    Pr_{LOS} = \exp{\left(-\frac{r_{2D}}{k_s}\right)},
\end{equation}
where $k_s$ given by
\begin{equation}
 {k_s} = 
\begin{cases}
-\frac{r_{c}}{\ln{(1-\varsigma)}}, \ $for InF-SL and InF-DL$, \\
-\frac{r_{c}}{\ln{(1-\varsigma)}}\cdot\frac{h_{BS}-h_{MTD}}{h_c- h_{MTD}}, \ $for InF-SH and InF-DH$,
\end{cases}
\end{equation}
where $r_c$ is the clutter size, $\varsigma$ is the clutter density, and $h_c$ is the effective clutter height. We train our model for all four types of InF scenarios to detect the active MTDs. InF environment parameters are presented in Table~\ref{tab2}.
Here, we define different $r_{2D}$ ranges for all InF scenarios to compare the performances in specific SNR values while maintaining the channel model condition that, $1 \leq r_{3D} \leq 600\ m$ given in~\cite{3gpp.138.901}. 
\begin{table*}[tb]
\caption{InF environment parameters}
\begin{center}
\begin{tabular}{llllllllll}
\hline
InF scenario& $h_{BS}$ & $h_{MTD}$ & $\varsigma$ & $h_c$ & $r_c$ & $r_{2D} range$ & $a$ & $b$ & $\sigma_{NLOS}$\\
\hline
InF-SL & $1.5\ m$ & $1.5\ m$ & $20 \%$ & $2\ m$ & $10\ m$ & $25 - 200\ m$ & $33.00$ & $25.50$ & $5.70$\\
InF-DL & $1.5\ m$ & $1.5\ m$ & $60 \%$ & $6\ m$ & $2\ m$ & $18 - 108\ m$ & $18.60$ & $35.70$ & $7.20$\\
InF-SH & $8.0\ m$ & $1.5\ m$ & $20 \%$ & $2\ m$ & $10\ m$ & $40 - 488\ m$ & $32.40$ & $23.00$ & $5.90$\\
InF-DH & $8.0\ m$ & $1.5\ m$ & $60 \%$ & $6\ m$ & $2\ m$ & $24 - 420\ m$ & $33.63$ & $21.90$ & $4.00$\\
\hline
\end{tabular}
\label{tab2}
\end{center}
\end{table*}
Furthermore, we used the same ORs, $N$ in both SCMA and MUSA, $p_i$, noise spectral density, noise figure at BS, and DNN parameters same as the Table~\ref{simulation1}. In addition, we consider $f_c = 28$\ GHz and transmission bandwidth $100$\ MHz for the mmWave InF simulations.

\begin{figure}
     \centering
     \begin{subfigure}[b]{0.46\linewidth}
         \centering
         \includegraphics[width=\linewidth]{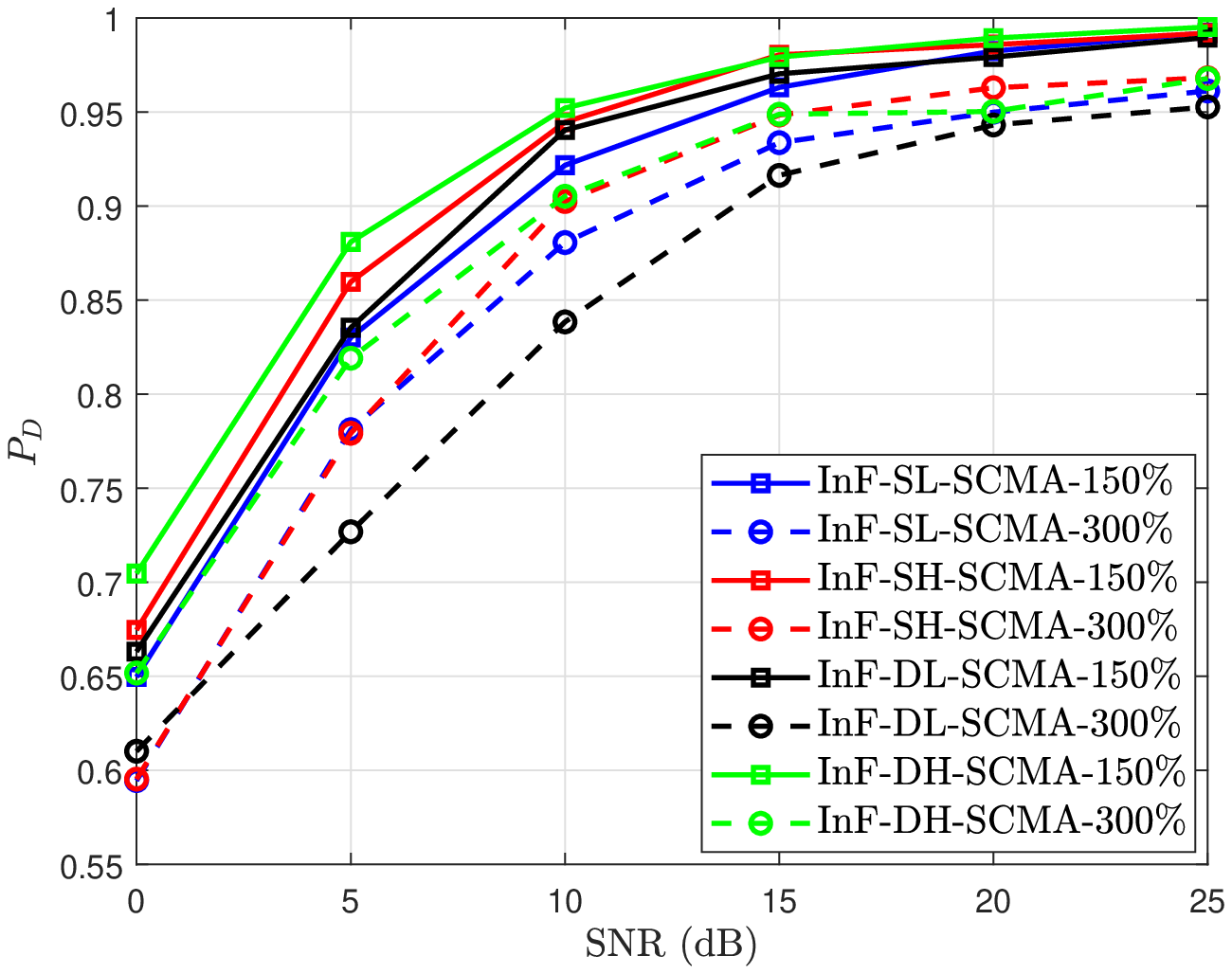}
         \caption{SCMA}
         \label{fig:mmWave_SCMA}
     \end{subfigure}
     \hspace{-1em}
     \begin{subfigure}[b]{0.46\linewidth}
         \centering
         \includegraphics[width=\linewidth]{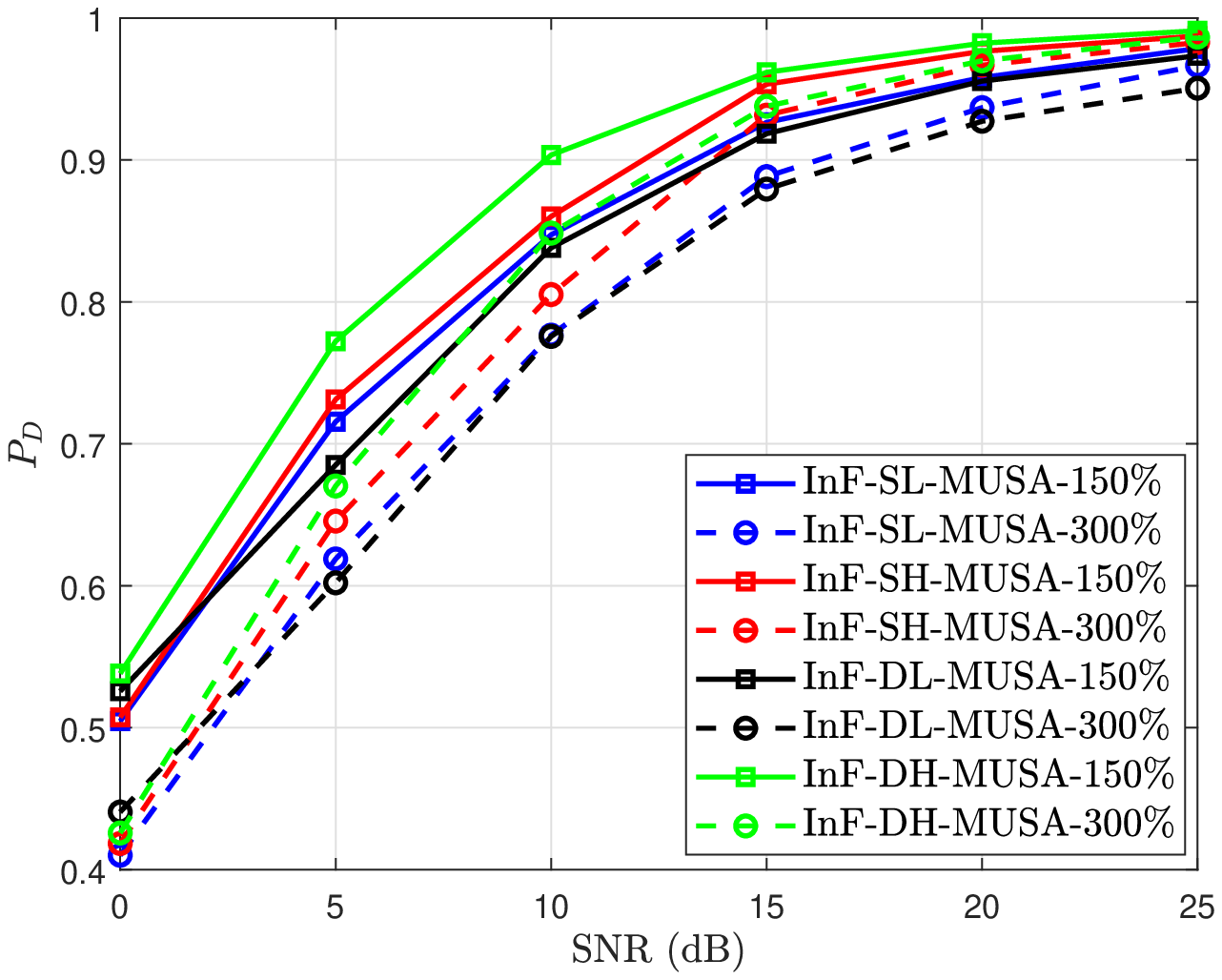}
         \caption{MUSA}
         \label{fig:mmWave_MUSA}
     \end{subfigure}
        \caption{Probability of detection (recall) versus the SNR for two different overloading ratios for (a) SCMA codebooks ($N$ = 90, $n$ = 3) and (b) MUSA spreading sequences ($N$ = 21, $n$ = 2) with four different number of indoor factory environments.}
        \label{fig:two_graphs5}
\end{figure}

Finally, we simulate the four different InF environments and evaluate the performance of the DNN MUD for the SMV system to verify the algorithm's applicability in future smart factories. Fig.~\ref{fig:two_graphs5} presents the probability of detection versus SNR for InF-SL, InF-SH, InF-DL, and InF-DH. Here, we observe that InF-DH achieves the highest overall performance in both SCMA and MUSA scenarios because of the higher chance of the LOS probability. Also, due to the $\varsigma$ difference between InF-DH and InF-SH, InF-SH reaches the second-highest overall performance. Likewise, the performance of InF-DL and InF-SL follows the argument as mentioned earlier. 
\section{Conclusion}
In this paper, we have considered jointly detecting the sparsity of the received signal and corresponding active devices in two different grant-free NOMA mMTC systems in the absence of CSI at the BS. The primary motivation is to blindly detect active devices in SMV and MMV systems with diverse ORs and explore the performance in a mmWave system setting with lower computational complexity. We have proposed a pre-activated ResNet-based DNN-MUD architecture to address the challenging sparse signal recovery problem. For the detailed examination of the proposed DNN-MUD, we have included an SCMA codebook design for a large number of devices and MUSA spreading sequence generation with a scalable heuristic algorithm to determine the spreading sequences. Furthermore, we have presented complete architecture details, including pre-activation, hidden layer design, post-activation, dropout, regularization, choice of the loss function, training data generation, and label annotation. In numerical simulations, we have carried out the system performance of the DNN-MUD using a calibration curve to validate the calibration of the proposed architecture. The performance of the proposed DNN-MUD scheme is comprehensively analyzed and is found to outperform well-known existing MUD algorithms. Simulation results of the SMV have shown that using DNN-MUD to detect the active MTD is a perfect choice even when the device activity level is high in a single time frame. Moreover, MMV systems have shown a very high probability of detection performance regardless of the codes' OR. The low computational complexity of the DNN-MUD further highlights the significance of the proposed approach, which encourages the less complex DNN-powered architectures for future mMTC applications. Device mobility, correlated systems where the device activity is mutually exclusive, and dynamic environments where the number of devices frequently varies in the system are some of the future extensions of this work that could be investigated.


%


\color{black}
\appendices
\section{Proof of Lemma~\ref{lemma_1}}
\begin{proof}
Consider a scenario for MTDs $a$ and $b$ are active (here, $b > a$), where we define the event $E_{a,b}$. We focus on bounding the probability in cases where not all label-pair interactions are covered, given as 
\begin{equation}
    P_s = 1 - Pr\bigg(\bigcup_{\forall a,b; b > a}E_{a,b}\bigg). 
\end{equation}
We use Dawson-Sankoff~\cite{Dawson} bound and investigation on~\cite{Rokach}, inequality for union of \begin{math} E_{a,b}, \forall a, b; b > a\end{math} is given by 
\begin{equation}
    Pr\bigg(\bigcup_{\forall a,b; b > a}E_{a,b}\bigg) \geq 2\frac{\bigg((\delta - 1)S_1 - S_2\bigg)}{\delta(\delta - 1)},
\end{equation}
where $S_1$ is the probability of specific pair of labels as 
\begin{equation}
    S_1 = \sum_{\forall a, b; b > a} Pr(E_{a,b}),
\end{equation}
$S_2$ is the combination of pair of labels as 
\begin{equation}
    S_2 = \sum_{\substack{\forall a, b, c, d; b > a, d > c,\\ a \neq c, b \neq d}} Pr(E_{a,b} \cap E_{c,d}), 
\end{equation}
\begin{equation}
    \delta = 2 + \left \lfloor \frac{2S_2}{S_1} \right \rfloor.
\end{equation}
Regarding $S_1$, there are $\Comb{N}{2}$ pairs of combinations; and we expect to select $\alpha$ subsets, where probability of having MTD $a$ and $b$ in a single label~\cite{Rokach} as 
\begin{equation}
    Pr(\lambda_a, \lambda_b) = \frac{\big(\Comb{N-2}{n} + 2\Comb{N-2}{n-1}\big)}{\Comb{N}{n}}. 
\end{equation}
Therefore, $S_1$ calculated as
\begin{equation}
    S_1 = \Comb{N}{2}\big( Pr(\lambda_a, \lambda_b) \big)^\alpha. 
\end{equation}
$S_2$ includes pairs of mutually exclusive ($S_{2,1}$) and mutually non-exclusive ($S_{2,2}$). First, we consider mutually exclusive part ($a \neq b, b \neq c, a \neq d, b \neq d$), which contain $3\Comb{N}{4}$ combinations and $\alpha$ subsets. We define the probability of MTDs $a, b, c, d$ are not included in the labelset as 
\begin{equation}
    Pr\big((\lambda_a, \lambda_b) \cap (\lambda_c, \lambda_d)\big) = \frac{\big(\Comb{N-4}{n} + 4\Comb{N-4}{n-1} + 4\Comb{N-4}{n-2}\big)}{\Comb{N}{n}}. 
    \label{s21}
\end{equation}
Therefore, $S_{2,1}$ calculated as
\begin{equation}
    S_{2,1} = 3\Comb{N}{4}\Big( Pr\big((\lambda_a, \lambda_b) \cap (\lambda_c, \lambda_d)\big) \Big)^\alpha. 
\end{equation}
Likewise, there are $3\Comb{N}{3}$ combinations, and define (\ref{s21}) for not mutually exclusive scenario as
\begin{equation}
    Pr\big((\lambda_a, \lambda_b) \cap (\lambda_c, \lambda_d)\big) = \frac{\big(\Comb{N-3}{n} + 3\Comb{N-3}{n-1} + \Comb{N-3}{n-2}\big)}{\Comb{N}{n}}. 
    \label{s22}
\end{equation}
Therefore, $S_{2,2}$ calculated as
\begin{equation}
    S_{2,2} = 3\Comb{N}{3}\Big( Pr\big((\lambda_a, \lambda_b) \cap (\lambda_c, \lambda_d)\big) \Big)^\alpha. 
\end{equation}
Hence, $S_2$ calculated as addition of $S_{2,1}$ and $S_{2,2}$. 
\end{proof}
\section{Proof of Theorem~\ref{theorem_1}}
\begin{proof}
There exists a sufficient amount of label combinations $\alpha$ to state the firm convergence of the problem in (\ref{reform}). Here, $D = \kappa \times \alpha$, where $\kappa$ is the number of channel realizations. Even increasing $\kappa$ does not necessarily guarantee $P_D$, specifically in implementation; however, $\alpha$ influences $P_D$, $P_M$, and the training time of the model. A minimum number of  $\alpha$ eases the training yet leads to an increase of $P_M$; however, the training data drastically increases when $\alpha \rightarrow \Comb{N}{n}$. 

We use the Lemma~\ref{lemma_1} to show that $P_s \rightarrow 1$, when $\alpha$ increase. When $P_s$ reaches $1$, we can confirm that the sufficient condition of the convergence is satisfied. In addition, we can find a suitable $\kappa$ when \begin{math} \mathcal{L}({p_j},\hat{p_j}) \rightarrow 0 \end{math} based on (\ref{loss}). Furthermore, based on Lemma~\ref{lemma_1}, the massive number of $N$ and range of $n$ also influence the convergence, which cannot be avoided.  
\end{proof}

\color{black}



%

\bibliographystyle{IEEEtran}
\bibliography{main.bib}

%








\end{document}